\documentclass[11pt]{article}
\usepackage{cancel}
\usepackage[pdftex]{color}
\usepackage{graphicx}
  \usepackage{amsthm,amsfonts}
  \usepackage{amsmath}
\usepackage{slashed}
\usepackage{ulem}
\usepackage{cleveref}
\usepackage{cite}
\newcommand{\bea}   {\begin{eqnarray}}
\newcommand{\eea}   {\end{eqnarray}}

\def\zzg{${\mathbb Z}_2\times{\mathbb Z}_2$-graded }

\def\ztwo{{\mathbb{Z}}_2}

\def\zthree{{\mathbb{Z}}_3}
\def\zthreetwo{{\mathbb{Z}}_3\times{\mathbb Z}_3}

\begin{document}
\renewcommand{\thefootnote}{\fnsymbol{footnote}}

\thispagestyle{empty}

\title{Braided {{quantum mechanics}} and Majorana qubits at third root \\of unity: 
 a color Heisenberg-Lie (super)algebra framework }

\author{Zhanna Kuznetsova\thanks{{E-mail: {\it zhanna.kuznetsova@ufabc.edu.br}}}\quad and\quad
Francesco Toppan\thanks{{E-mail: {\it toppan@cbpf.br}}}
\\
\\
}
\maketitle

\centerline{$^{\ast}$ {\it UFABC, Av. dos Estados 5001, Bangu,}}\centerline{\it { cep 09210-580, Santo Andr\'e (SP), Brazil. }}
~\\
\centerline{$^{\dag}$
{\it CBPF, Rua Dr. Xavier Sigaud 150, Urca,}}
\centerline{\it{
cep 22290-180, Rio de Janeiro (RJ), Brazil.}}
~\\
\maketitle
\begin{abstract}
We introduce {\it color} Heisenberg-Lie (super)algebras graded by the abelian groups  $\zthree^2$, $\ztwo^p\times\zthree^2$ for $p=1,2,3$, and investigate the properties of their associated multi-particle quantum  
paraoscillators.\par In the Rittenberg-Wyler's color Lie (super)algebras framework the above abelian groups are the simplest ones which induce {\it mixed brackets} interpolating commutators and anticommutators. These mixed brackets allow to accommodate two types of parastatistics: one based on the permutation group (beyond bosons and fermions in any space dimension) and an anyonic parastatistics based on the braid group. In both such cases the two broad classes of paraparticles are given by parabosons and parafermions.\par Mixed-bracket parafermions are created by nilpotent operators; they satisfy a generalized Pauli exclusion principle leading to roots-of-unity truncations in their multi-particle energy spectrum (braided Majorana qubits and their Gentile-type parastatistics are recovered in this color Lie superalgebra setting). \par Mixed-bracket parabosons do not admit truncations of the spectrum;  the minimal detectable signature of their parastatistics is encoded in the measurable probability density of two indistinguishable parabosonic oscillators in a given energy eigenstate. 
\end{abstract}
\vfill

\rightline{CBPF-NF-003/25}

\newpage

\section{Introduction}

Ordinary Lie (super)algebras were extended by Rittenberg and Wyler in \cite{{rw1},{rw2}} to ``{\it Color Lie (super)algebras}" defined in terms of a general abelian-group grading. Quite sooon this new mathematical structure was  systematically analyzed by Scheunert in \cite{sch}. \par
The simplest examples of this class of theories are obtained for gradings based on the ${\mathbb Z}_2^n$ groups. For these gradings the defining brackets are ordinary commutators/anticommutators organized in peculiar ways. They imply a class of parastatistics which can be referred to, see \cite{nbits}, as ``$n$-$bit$" parastatistics. The resulting paraparticles are exchanged under the permutation group and can exist in any space dimension.\par
Beyond the ${\mathbb Z}_2^n$ groups, new possibilities are opened. Already in the original \cite{rw1} paper, Rittenberg-Wyler produced, for the $\zthreetwo:= {\mathbb Z}_3^2$ group, a color Lie algebra whose defining graded brackets consist of nontrivial linear combinations of commutators/anticommutators (from now on we will apply the term {\it mixed-bracket} to such nontrivial linear combinations).\par
At the beginning, some limited attention was given to the physical applications of ${\mathbb Z}_2^2$-graded color Lie (super)algebras, see e.g. \cite{{vas},{jyw}}; on the other hand, the recent decade experienced a boom (more on that later) of  papers devoted to mathematical investigations and physical applications of  ${\mathbb Z}_2^n$-graded color Lie (super)algebras.\par
What about the more general case, namely the possible physical applications offered by {\it mixed-bracket} color Lie (super)algebras? Strangely enough, almost nothing has been done. Perhaps the most notable exception consists in the ${\mathbb Z}_4\times{\mathbb Z}_4$-graded extensions of the Poincar\'e algebra introduced in \cite{{wito},{wito2}}.\par
In this paper we introduce the simplest  mixed-bracket {\it color} extensions of the Heisenberg-Lie (super)algebras (their grading abelian groups are given by $\zthree^2$ and $\ztwo^p\times\zthree^2$ for $p=1,2,3$). We construct their associated multi-particle quantum (para)oscillators and investigate their physical properties.
The two broad classes of mixed-bracket paraparticles are given by parabosons and parafermions. \par Mixed-bracket parafermions are created by nilpotent operators and satisfy a generalized Pauli exclusion principle which implies truncations in their multi-particle energy spectrum. Their simplest physical application consists in reformulating the multi-particle braided Majorana qubits \cite{{topqubits},{topvoli}} in the color Lie superalgebra framework; the truncations  lead to an implementation of a Gentile-type parastatistics \cite{gen}. \par 
Mixed-bracket parabosons do not admit truncations of the spectrum. We show that the minimal detectable signature of their parastatistics is encoded in the measurable probability density of two indistinguishable parabosonic oscillators in a given energy eigenstate. \par
The structure of the paper is the following.
The paper is dividided in two parts separated by an {\it Intermezzo} (Section {\bf 6}). \par
The first part (Sections {\bf 2}-{\bf 5}) details the foundation of the theory. Section {\bf 2} reviews, following \cite{sch}, the basic notions of color Lie (super)algebras. Section {\bf 3} presents several ${\mathbb Z}_3$-graded matrices used as building blocks in the construction of the color Lie (super)algebras later introduced. Section {\bf 4} presents the two inequivalent $\zthreetwo$-graded color Lie algebras and the  four inequivalent $\ztwo\times\zthreetwo$ color Lie (super)algebras. Graded para-oscillators (both parabosonic and parafermionic) are introduced in Section {\bf 5}.\par
The second part (Sections {\bf 7,8}) describes applications. Section {\bf 7} points out which is the minimal signature for mixed-bracket indistinguishable parabosons: it is given by multi-particle probability densities in certain energy eigenstates. 
In Section {\bf 8} it is shown that a signature of mixed-bracket indistinguishable parafermions is given by truncations of the energy spectra (these truncations generalize the Pauli exclusion principle of ordinary fermions). The connection with  braided Majorana qubits (at the $s=3,6$ roots of unity levels) is made.\par
The Section {\bf 6} {\it Intermezzo} discusses two scenarios for mixed-bracket multi-particle sectors; the first scenario is based on graded Hopf algebras and their coproducts, while the second one is based on the symmetrization of the creation operators.  It is further pointed out that mixed-bracket color Lie (super)algebras can be applied to two types of parastatistics (both the parastatistics associated with the permutation group and the anyonic parastatistics associated with the braid group).\par
The paper is complemented by three Appendices.\par
{{To put our paper in due context, Appendix {\bf A} presents a state-of-the-art-account of color Lie (super)algebras and  their physical applications; some mathematical structures related to the present work are also discussed.}}\par
The important notion of {\it  roots-of-unity level} is defined in Appendix {\bf B}.\par
{{For selfconsistency of the paper, the notion of Gentile parastatistics, braided Majorana qubits and their properties, are briefly summarized in the Appendix {\bf C}.}}\par
Several comments are presented in the {\bf Conclusions}. Besides summarizing the main results of the paper,
we present a list of further investigations and main open questions concerning physical applications of mixed-bracket color Lie (super)algebras and their relation with other mathematical structures.

\section{Review of color Lie (super)algebras}

For self-consistency of the paper we introduce definitions and basic features of color Lie (super)algebras. We follow Scheunert's presentation \cite{sch} which is more convenient for our purposes than Rittenberg-Wyler's \cite{{rw1},{rw2}} approach. On the other hand, since it is more widely accepted in the physical literature, we refer to these structures by using the term ``{color Lie (super)algebras}" coined by Rittenberg-Wyler, rather than ``{$\varepsilon$ Lie algebras}" introduced by Scheunert.\par
Let $\Gamma$  be an Abelian group and $V$ a
$\Gamma$-graded vector space $V=\oplus_\gamma V_\gamma$, $\gamma\in \Gamma$, so that 
the homogeneous elements belong to $V_\gamma$.\par
A linear mapping $g: V\rightarrow W$ between $\Gamma$-graded spaces is homogeneous of degree $\gamma\in \Gamma$ if $g(V_\alpha)\subset W_{\alpha+\gamma}$.\par
For three $\Gamma$-graded spaces $U,V,W$ let $h: U\rightarrow V$, $g:V\rightarrow W$ be homogeneous of degree $deg(h) = \alpha$, $deg(g) =\beta$; it follows that the composition map $g\circ h: U\rightarrow W$ is homogeneous of degree $deg(g\circ h)= \alpha+\beta$.\par
An algebra $S$ is $\Gamma$-graded if $S=\oplus_{\gamma\in\Gamma} S_\gamma$ as a vector space and $S_\alpha S_\beta\subset S_{\alpha+\beta}$ for any $\alpha,\beta\in \Gamma$.\\
Let $T$ be another $\Gamma$-graded algebra. An $S\rightarrow T$ homomorphism of $\Gamma$-graded algebras is an homomorphism of $S$ to $T$ which is a homogeneous mapping of degree zero.\par
{\it Definition of commutation factor $\varepsilon$ on abelian group $\Gamma$}: 
it is a bilinear map $\varepsilon: \Gamma\times \Gamma \rightarrow C^\ast$ (taking values in the punctured complex plane
${\mathbb C}^\ast\equiv {\mathbb C}\backslash \{0\}$) satisfying, for any $\alpha,\beta,\gamma$ in $\Gamma$, the properties:
\bea\label{epsilonproperties}
&&~~{\textrm{{\it i}): \quad $\varepsilon(\alpha,\beta)\cdot \varepsilon(\beta,\alpha) =1$,}}\nonumber\\
&&~{\textrm{{\it ii}): \quad $\varepsilon(\alpha,\beta+\gamma) = \varepsilon(\alpha,\beta)\cdot \varepsilon(\alpha,\gamma)$,}}\nonumber\\
&&{\textrm{{\it iii}): \quad $\varepsilon(\alpha+\beta,\gamma) =\varepsilon(\alpha,\gamma)\cdot\varepsilon(\beta,\gamma)$.}}
\eea
A $\Gamma$-graded $\varepsilon$-Lie algebra $L=\oplus_\gamma L_\gamma$ ({\it color Lie (super)algebra} in \cite{{rw1},{rw2}}) is defined by the brackets $\langle .,.\rangle: L\times L\rightarrow L$ which satisfy the conditions, for any $A,B,C\in L$ and respective $\alpha, \beta, \gamma$ gradings:
\bea\label{epsilonalgebras}
&&{\textrm{{\it i}) $\quad \langle A, B\rangle = -\varepsilon (\alpha,\beta)\langle B, A\rangle\qquad \quad\quad  $ (the $\varepsilon$-skew symmetry) and}}\nonumber\\
&&{\textrm{{\it ii})  $\quad \varepsilon(\gamma,\alpha)\langle A,\langle B,C\rangle\rangle + cyclic =0\quad $ (the $\varepsilon$-Jacobi identities).}}
\eea
It follows,  from the first relation in (\ref{epsilonproperties}), that $\varepsilon(\alpha,\alpha) =\pm1$. Motivated by physical applications,  Rittenberg-Wyler introduced the term {\it color Lie algebra}  if the sign $+1$ is valid for any $\alpha\in \Gamma$; if there exists at least one $\alpha \in\Gamma$ such that
$\varepsilon(\alpha,\alpha) =-1$, the corresponding algebra is called {\it color Lie superalgebra}.\par
For any associative $\Gamma$-graded space $S$ the introduction of a bracket $\langle.,.\rangle$ defined by
\bea\label{gradedbracket}
\langle A, B\rangle := AB-\varepsilon(\alpha,\beta) BA
\eea
induces, on the graded vector space $S$, a color Lie (super)algebra satisfying the equations (\ref{epsilonalgebras}).
\par
{\it Remark 1}: at the special value $\varepsilon(\alpha,\beta)=+1$ the graded bracket
$\langle A, B\rangle \equiv AB-BA=[A,B]$ defines an ordinary commutator while,
at  the special value $\varepsilon(\alpha,\beta)=-1$, the graded bracket
$\langle A, B\rangle \equiv AB+BA=\{A,B\}$ defines an anticommutator.\par
{\it Remark 2}: in this work the abelian groups $\Gamma$ under consideration belong to the (direct product of) additive groups ${\mathbb Z}_n$ of integers modulo $n$.  For special cases, such as the group  ${\mathbb Z}_2^p:= 
{\mathbb Z}_2\times \ldots \times {\mathbb Z}_2$ (taken $p$ times), all graded brackets are either commutators or anticommutators  (namely, $\varepsilon (\alpha,\beta)=\pm 1$ for any $\alpha,\beta\in {\mathbb Z}_2^p$). Only commutators/anticommutators are also encountered for the direct product group ${\mathbb Z}_2\times {\mathbb Z}_3$, while ${\mathbb Z}_3$ admits only commutators (one gets, $\forall \alpha,\beta\in {\mathbb Z}_3$, $\varepsilon (\alpha,\beta)=+1$). The group ${\mathbb Z}_3\times {\mathbb Z}_3$ provides, see \cite{rw1}, the simplest nontrivial example of consistent graded brackets which satisfy the (\ref{epsilonproperties}) properties and do not reduce to (anti)commutators (this means that $\varepsilon(\alpha,\beta)\neq \pm 1$ for at least a pair of $\alpha,\beta$ values). 
These types of brackets are the focus of the present work. We consider, in particular, the direct product groups
${\mathbb Z}_2^p\times {\mathbb Z}_3^q$ for integer values $p, q =0,1,2,\ldots $). \par
Throughout the paper the ${\mathbb Z}_2$-grading will be denoted, following \cite{nbits}, by one bit ($0,1$), while the ${\mathbb Z}_3$-grading will be denoted by a trit (${\underline 0},{\underline 1}, {\underline 2}$). The integers in a trit are underlined to avoid confusion with the $0,1$ values entering the ${\mathbb Z}_2$-grading.

\subsection{Iterated commutation factors}

We present a construction which induces a consistent subclass of commutation factors (satisfying the (\ref{epsilonproperties}) properties) for the direct {{product groups}} ${\mathbb Z}_2\times \Gamma$ and ${\mathbb Z}_3\times \Gamma$ once a choice of $\varepsilon$ commutation factors of an $n$-dimensional abelian group $\Gamma$ is given. Let $U$ be a table presenting the commutation factors of $\Gamma$ in a $n\times n$ array (throughout the paper the $\varepsilon(\alpha,\beta)$ commutation factor is expressed as the entry associated with the $\alpha$ row and $\beta$ column).  Then, commutation factors are obtained for:\\
~\\
{\it a}) {\it the ${\mathbb Z}_2\times \Gamma$ abelian group}: two sets of commutation factors are obtained for $\delta=\pm 1$. They are presented in a $2n\times 2n $ array according to the scheme
\bea\label{aiter}
|U| &\Rightarrow &\begin{array}{|cc|}U&U\\U&\delta U\end{array}~;
\eea
~\\
{\it b}) {\it the ${\mathbb Z}_3\times \Gamma$ abelian group}: a set of commutation factors is given by the following $3n\times 3n $ array
\bea\label{biter}
|U| &\Rightarrow &\begin{array}{|ccc|}U&U&U\\U& U&U\\U&U&U\end{array}~.
\eea
\\~\\
As an example of $a$), an ordinary Lie algebra with commutation factor $1$ produces
\bea
{\textrm{for $\delta=+1$}}&:& \quad 
|1| \Rightarrow \begin{array}{|cc|}1&1\\1&1\end{array}~,~~~ {\textrm{i.e., an ordinary Lie algebra endowed with a ${\mathbb Z}_2$ grading}},\nonumber\\
{\textrm{for $\delta=-1$}}&:& \quad 
|1| \Rightarrow \begin{array}{|cr|}1&1\\1&-1\end{array}~,~ {\textrm{i.e., an ordinary Lie superalgebra.}}
\eea
\\~\\
As an example of $b$), an ordinary Lie algebra with commutation factor $1$ produces
\bea
|1| &\Rightarrow &\begin{array}{|ccc|}1&1&1\\1& 1&1\\1&1&1\end{array}~,~~~ {\textrm{i.e., an ordinary Lie algebra endowed with a ${\mathbb Z}_3$ grading}}.
\eea
The above one  is the only consistent ${\mathbb Z}_3$-graded color Lie (super)algebra (which turns out to be a trivial ordinary Lie algebra). By inserting the gradings of the rows/columns, the above $3\times 3$ array reads as
\bea
&\begin{array}{|c|ccc|}\hline &{\underline 0}&{\underline 1}&{\underline 2}\\ \hline {\underline 0}& 1&1&1\\{\underline 1}&1&1&1\\ {\underline 2}&1&1&1 \\ \hline\end{array}
\eea
\\
~\\
By applying the {\it a}) iteration to the above commutation factors we obtain the only consistent ${\mathbb Z}_2\times {\mathbb Z}_3$-graded color Lie (super)algebras, expressed by the $\delta$-dependent $6\times 6$ arrays
\bea
&\begin{array}{|c|ccc|ccc|}\hline &0{\underline{0}}&0{\underline{ 1}}&0{\underline{2}}&1{\underline{0}}&1{\underline{1}}&1{\underline{2}}  \\ \hline 
0{\underline {0}}& 1&1&1&1&1&1\\
0{\underline{ 1}}&1&1&1&1&1&1 \\    
0{\underline {2}}& 1&1&1&1&1&1\\ \hline
1{\underline{ 0}}&1&1&1&\delta&\delta&\delta \\    
1{\underline {1}}& 1&1&1&\delta&\delta&\delta\\
1{\underline{2}}&1&1&1&\delta&\delta&\delta \\    \hline\end{array}
\eea
The rows/columns are expressed in terms of the  $0,1$ bit notation for ${\mathbb Z}_2$ and the ${\underline 0}, {\underline 1}, {\underline 2}$ trit notation
 for ${\mathbb Z}_3$.\par
The two inequivalent cases are recovered for:\par
${\delta=+1}$, producing an ordinary Lie algebra and\par
$\delta =-1$, producing an ordinary Lie superalgebra.

\section{Building blocks: $\zthree$-graded matrices}

In the construction of ${\mathbb Z}_3\times{\mathbb Z}_3$-graded color Lie algebras we use, as building blocks, ${\mathbb Z}_3$-graded $3\times 3$ matrices. We present here their main features and introduce some sets of $\zthree$-graded matrices which are relevant for this work.\par
The non-vanishing entries of ${\mathbb Z}_3$-graded $3\times 3$ matrices ${\bf M}_{\underline k}$, with
 ${\underline k}= {\underline 0},{\underline 1}, {\underline 2}$, are 
\bea
&{\bf M}_{\underline 0}=\left(\begin{array}{ccc}\ast&0&0\\0&\ast&0\\0&0&\ast\end{array}\right),\qquad {\bf M}_{\underline 1}=\left(\begin{array}{cccc}0&\ast&0\\0&0&\ast\\ \ast&0&0\end{array}\right),\qquad {\bf M}_{\underline 2}=\left(\begin{array}{cccc}0&0&\ast\\\ast&0&0\\0&\ast&0\end{array}\right).&
\eea 
Under matrix multiplication ${\bf M}_{\underline k}\cdot {\bf M}_{\underline k'}$ has grading $ {\underline{k+k'}}= {\underline k}+{\underline k'} ~ mod ~3$.
\\
~\\
The following sets of ${\mathbb Z}_3$-graded $3\times 3$ matrices are used in this paper. Their entries are $0, 1, j, j^2$, where $j$ is a third root of unity satisfying $j^3=1$. We have:\\
~\\
{\it i}) the ${\underline 0}$-graded diagonal matrices
\bea
&{N}_{\underline 0}=\left(\begin{array}{ccc}1&0&0\\0&1&0\\0&0&1\end{array}\right),\qquad {N}_{\underline 0}'=\left(\begin{array}{cccc}1&0&0\\0&j&0\\ 0&0&j^2\end{array}\right),\qquad {N}_{\underline 0}''=\left(\begin{array}{cccc}1&0&0\\0&j^2&0\\0&0&j\end{array}\right);&\label{z3one}
\eea 
~\\
{\it ii}) the commuting matrices
\bea
&{N}_{\underline 0}=\left(\begin{array}{ccc}1&0&0\\0&1&0\\0&0&1\end{array}\right),\qquad {N}_{\underline 1}=\left(\begin{array}{cccc}0&1&0\\0&0&1\\ 1&0&0\end{array}\right),\qquad {N}_{\underline 2}=\left(\begin{array}{cccc}0&0&1\\1&0&0\\0&1&0\end{array}\right),&\label{z3two}
\eea 
satisfying, for any ${\underline k}, {\underline k'}$, $N_{\underline k}N_{\underline k'}=N_{\underline k'}N_{\underline k}$;\\
~\\
{\it iii}) the ${\underline 1}$-graded matrices, 
\bea
&{Q}_{+1}=\left(\begin{array}{ccc}0&1&0\\0&0&j\\j^2&0&0\end{array}\right),\qquad {Q}_{+2}=\left(\begin{array}{cccc}0&j&0\\0&0&1\\ j^2&0&0\end{array}\right),\qquad {Q}_{+3}=\left(\begin{array}{cccc}0&1&0\\0&0&1\\1&0&0\end{array}\right)&\label{z3three}
\eea 
and their hermitian conjugates $Q_{+i} ^\dagger := Q_{-i}$, given by\\
~\\
{\it iv}) the ${\underline 2}$-graded matrices
\bea
&{Q}_{-1}=\left(\begin{array}{ccc}0&0&j\\1&0&0\\0&j^2&0\end{array}\right),\qquad {Q}_{-2}=\left(\begin{array}{cccc}0&0&j\\j^2&0&0\\ 0&1&0\end{array}\right),\qquad {Q}_{-3}=\left(\begin{array}{cccc}0&0&1\\1&0&0\\0&1&0\end{array}\right).&\label{z3four}
\eea 

\section{Some relevant color Lie (super)algebras}

For an abelian group $\Gamma$ of small dimension $n$ it is computationally feasible to impose the (\ref{epsilonproperties}) properties on an $n\times n$ array, resulting in the whole set of inequivalent, admissible commutation factors. We present here the results for $\zthreetwo$
 and $\ztwo\times \zthreetwo$. All  resulting commutation factors are given by roots of unity. For $\zthreetwo$ we recover the color Lie algebra introduced in \cite{rw1}.

\subsection{The $\zthreetwo$-graded color Lie algebra}

The consistent commutation factors of the ${\mathbb  Z}_3\times {\mathbb Z}_3$-graded abelian group are presented as entries of the following $9\times 9$ array. The entries are given by the numbers $1,j, j^2$, where $j$ is a third root of unity satisfying $j^3=1$.  The rows and columns are labeled by the two-trit notation. We have
\bea
&\begin{array}{|c|ccc|ccc|ccc|}\hline &{\underline{00}}&{\underline{ 01}}&{\underline{02}}&{\underline{10}}&{\underline{11}}&{\underline{12}} &{\underline{20}}&{\underline{21}}&{\underline{ 22}} \\ \hline 
{\underline {00}}& 1&1&1&1&1&1&1&1&1\\
{\underline{ 01}}&1&1&1&j^2&j^2&j^2&j&j&j \\  
{\underline {02}}& 1&1&1&j&j&j&j^2&j^2&j^2\\ \hline
{\underline {10}}&1&j&j^2&1&j&j^2&1&j&j^2 \\ 
{\underline {11}}& 1&j&j^2&j^2&1&j&j&j^2&1\\
{\underline {12}}&1&j&j^2&j&j^2&1&j^2&1&j \\  \hline
{\underline {20}}& 1&j^2&j&1&j^2&j&1&j^2&j\\
{\underline {21}}&1&j^2&j&j^2&j&1&j&1&j^2 \\ 
{\underline {22}}& 1&j^2&j&j&1&j^2&j^2&j&1\\  \hline\end{array}&\label{z3z3array}
\eea
Three sets of commutation factors are recovered for each one of the three solutions of $j^3=1$, given by
$j_1 = e^{\frac{2\pi i}{3}}, ~j_2 =e^{\frac{4\pi i}{3}}, ~ j_3 =1$. \par
As discussed in Appendix {\bf B}, 
$j_1,j_2$ are level-$3$ roots of unity which, in particular, satisfy the equations
$1+j_1+j_1^2=1+j_2+j_2^2=0$. On the other hand $j_3$ is a level-$1$ root of unity which satisfies $1+j_3+j_3^2=3$. Setting {{$j=j_3=1$}} in the above table produces an ordinary Lie algebra, while setting $j_1,j_2$  produces a non-trivial color Lie algebra whose (\ref{gradedbracket}) graded brackets are not all given by
commutators/anticommutators. \par
The ${\mathbb  Z}_3\times {\mathbb Z}_3$-graded color Lie algebras induced by $j_1, j_2$ are isomorphic. 
This results from the fact that the $j_2$ array in (\ref{z3z3array}) can be recovered from the $j_1$ array under the
${\underline {k1}}\leftrightarrow {\underline{k2}}$ permutation (for any 
${\underline k}= {\underline 0},{\underline 1},{\underline 2}$) of the graded sectors entering the rows/columns.\par
There are two inequivalent choices for consistent $\zthreetwo$-graded commutation factors. We have
\bea\label{z3z3cor}
~~~i: && j=j_3=+1, \quad {\textrm{producing an ordinary Lie algebra and}}\nonumber\\~~
~ii: &&j= j_1, \quad ~~\qquad {\textrm{giving a nontrivial color Lie algebra.}}
\eea
{\it Remark}: The computation of all sets of admissible commutation factors (the $\varepsilon({{\underline{ij}},{\underline {kl}}})$ entries in the (\ref{z3z3array}) array) is straightforward. The first equation in (\ref{epsilonproperties}) implies $9$ diagonal plus $36$ upper triangular independent parameters. They are constrained by the two remaining equations in (\ref{epsilonproperties}). By setting $\alpha =  {\underline{00}}$  and $\gamma =  {\underline{ij}}$ we get, from the third equation,
$\varepsilon({\underline{00}},{\underline{ij}})=1$ for any choice of ${\underline{i}},   {\underline{j}}$. The other parameters  are determined by different relations. 
For instance $\varepsilon({\underline{12}},{\underline{ij}})= \varepsilon({\underline{21}},{\underline{ij}})^{-1}$ is obtained from from  $1=\varepsilon({\underline{00}},{\underline{ij}})=\varepsilon({\underline{12}},{\underline{ij}})\cdot \varepsilon({\underline{21}},{\underline{ij}})$. \\
The presence of the $j^3=1$ constraint is induced, e.g., from the series of relations $\varepsilon({\underline{02}},{\underline{11}})=\varepsilon({\underline{01}},{\underline{11}})\cdot \varepsilon({\underline{01}},{\underline{11}})$ and 
$1=\varepsilon({\underline{00}},{\underline{11}})=\varepsilon({\underline{02}},{\underline{11}})\cdot \varepsilon({\underline{01}},{\underline{11}})$ which imply  $\varepsilon({\underline{01}},{\underline{11}})^3=1$.\\
Implementing the complete set of relations produces the (\ref{z3z3array}) array.

\subsection{The $\ztwo\times\zthreetwo$ graded color Lie (super)algebras}

Direct computations give the whole class of inequivalent commutation factors, satisfying the (\ref{epsilonproperties}) properties, for the ${\mathbb Z}_2\times {\mathbb  Z}_3\times {\mathbb Z}_3$ grading. The results are summarized in the following $18\times 18$ table with rows/columns labeled by 1 bit and 2 trits:
{\footnotesize{
\bea
&\begin{array}{|c|ccc|ccc|ccc|ccc|ccc|ccc|}\hline &0{\underline{00}}&0{\underline{ 01}}&0{\underline{02}}&0{\underline{10}}&0{\underline{11}}&{0{\underline{12}}} &0{\underline{20}}&0{\underline{21}}&0{\underline{ 22}}
&1{\underline{00}}&1{\underline{ 01}}&1{\underline{02}}&1{\underline{10}}&1{\underline{11}}&1{\underline{12}} &1{\underline{20}}&1{\underline{21}}&1{\underline{ 22}} \\ \hline 
0{\underline {00}}& 1&1&1&1&1&1&1&1&1& 1&1&1&1&1&1&1&1&1\\
0{\underline {01}}&1&1&1&j^2&j^2&j^2&j&j&j&1&1&1&j^2&j^2&j^2&j&j&j \\  
0{\underline {02}}& 1&1&1&j&j&j&j^2&j^2&j^2& 1&1&1&j&j&j&j^2&j^2&j^2\\ \hline
0{\underline {10}}&1&j&j^2&1&j&j^2&1&j&j^2&1&j&j^2&1&j&j^2&1&j&j^2 \\ 
0{\underline {11}}& 1&j&j^2&j^2&1&j&j&j^2&1& 1&j&j^2&j^2&1&j&j&j^2&1\\
0{\underline {12}}&1&j&j^2&j&j^2&1&j^2&1&j &1&j&j^2&j&j^2&1&j^2&1&j\\  \hline
0{\underline {20}}& 1&j^2&j&1&j^2&j&1&j^2&j& 1&j^2&j&1&j^2&j&1&j^2&j\\
0{\underline {21}}&1&j^2&j&j^2&j&1&j&1&j^2&1&j^2&j&j^2&j&1&j&1&j^2  \\ 
0{\underline {22}}& 1&j^2&j&j&1&j^2&j^2&j&1& 1&j^2&j&j&1&j^2&j^2&j&1\\  \hline
1{\underline {00}}& 1&1&1&1&1&1&1&1&1& \delta&\delta&\delta&\delta&\delta&\delta&\delta&\delta&\delta\\
1{\underline {01}}&1&1&1&j^2&j^2&j^2&j&j&j&\delta&\delta&\delta&\delta j^2&\delta j^2&\delta j^2&\delta j&\delta j&\delta j \\  
1{\underline {02}}& 1&1&1&j&j&j&j^2&j^2&j^2& \delta&\delta&\delta&\delta j&\delta j&\delta j&\delta j^2&\delta j^2&\delta j^2\\ \hline
1{\underline {10}}&1&j&j^2&1&j&j^2&1&j&j^2&\delta&\delta j&\delta j^2&\delta &\delta j&\delta j^2&\delta&\delta j&\delta j^2 \\ 
1{\underline {11}}& 1&j&j^2&j^2&1&j&j&j^2&1& \delta&\delta j&\delta j^2&\delta j^2&\delta&\delta j&\delta j&\delta j^2&\delta\\
1{\underline {12}}&1&j&j^2&j&j^2&1&j^2&1&j &\delta&\delta j&\delta j^2&\delta j&\delta j^2&\delta&\delta j^2&\delta&\delta j\\  \hline
1{\underline {20}}& 1&j^2&j&1&j^2&j&1&j^2&j& \delta&\delta j^2&\delta j&\delta&\delta j^2&\delta j&\delta&\delta j^2&\delta j\\
1{\underline {21}}&1&j^2&j&j^2&j&1&j&1&j^2&\delta&\delta j^2&\delta j&\delta j^2&\delta j&\delta &\delta j&\delta&\delta j^2  \\ 
1{\underline {22}}& 1&j^2&j&j&1&j^2&j^2&j&1& \delta&\delta j^2&\delta j&\delta j&\delta &\delta j^2&\delta j^2&\delta j&\delta\\  \hline
\end{array}&\nonumber\\&&\label{z2z3z3array}
\eea}}
The above table is expressed in terms of a third root of unity $j$ satisfying $j^3=1$ and of a $\delta =\pm 1$ sign. It turns out that all inequivalent ${\mathbb Z}_2\times {\mathbb  Z}_3\times {\mathbb Z}_3$-graded commutation factors are recovered from the iteration (\ref{aiter}) applied to the (\ref{z3z3array}) array.\par
Setting $\delta=-1 $ in the above table induces, following the Rittenberg-Wyler's terminology, a {\it color Lie superalgebra}.  By repeating the previous Subsection analysis it results that, out of the $6=3\times 2$ arrays obtained by setting $j= j_1,j_2,j_3$ and $\delta =\pm 1$, four of them produce inequivalent commutation factors.\par
~\par
There are $4$ inequivalent ${\mathbb Z}_2\times {\mathbb  Z}_3\times {\mathbb Z}_3$-graded color Lie (super)algebras, given by
\bea\label{summary}
~~~i: && j_3=+1, \qquad ~~ \delta=+1\qquad {\textrm{(ordinary Lie algebra),}}\nonumber\\~~
~ii: && j_3=+1, \qquad ~~\delta=-1\qquad {\textrm{(ordinary Lie superalgebra),}}\nonumber\\
iii: && j_1=e^{\frac{2\pi i}{3}}, \qquad \delta=+1\qquad {\textrm{(color Lie algebra),}}\nonumber\\
~iv: && j_1=e^{\frac{2\pi i}{3}}, \qquad \delta=-1\qquad {\textrm{(color Lie superalgebra).}}
\eea

\subsection{A graded-abelian $\zthreetwo$ color Lie algebra}

The $\zthree$-graded $3\times 3$ matrices (\ref{z3one},\ref{z3two},\ref{z3three},\ref{z3four}) introduced in Section {\bf 3} allow, used as building blocks, to construct matrix representations of $\zthreetwo$-graded color Lie algebras. In particular, the following $9\times 9$ matrix representation of the graded-abelian $\zthreetwo$ color Lie algebra with one generator in each graded sector (for a total number of $9$ generators), is obtained. The matrices are denoted as $C_{\underline ij}$; the suffix indicates the $\zthreetwo$ grading. We have
\bea
&
&C_{\underline{00}} = N_{\underline 0}\otimes N_{\underline 0}, \qquad 
~~ C_{\underline{01}} = N_{\underline 0}'\otimes N_{\underline 1}, \qquad
~C_{\underline{02}} = N_{\underline 0}''\otimes N_{\underline 0}, \qquad\nonumber\\
&&C_{\underline{10}} = Q_{+1}\otimes N_{\underline 0}, \qquad 
C_{\underline{11}} = Q_{+2}\otimes N_{\underline 1}, \qquad
C_{\underline{12}} = Q_{+3}\otimes N_{\underline 2}, \qquad\nonumber\\
&&C_{\underline{20}} = Q_{-1}\otimes N_{\underline 0}, \qquad 
C_{\underline{21}} = Q_{-3}\otimes N_{\underline 1}, \qquad
C_{\underline{22}} = Q_{-2}\otimes N_{\underline 2}.\qquad\label{gradedcomm}
\eea
It is easily realized that the above matrices are graded-commutative. Indeed, their 
graded commutative brackets defined by 
\bea\label{z3z3brack}
\langle C_{\underline{ij}},C_{\underline{kl}}\rangle&:= &C_{\underline{ij}}C_{\underline{kl}}-\varepsilon({{\underline{ij}},{\underline {kl}}})C_{\underline{kl}}C_{\underline{ij}}
\eea
satisfy, for any choice of $\underline{i}, \underline{j},\underline{k},\underline{l}$, the vanishing relations
\bea
\langle C_{\underline{ij}},C_{\underline{kl}}\rangle
&=& 0.
\eea
The commutation factors $\varepsilon({{\underline{ij}},{\underline {kl}}})$ entering (\ref{z3z3brack}) are given in table ({\ref{z3z3array}).

\section{Graded para-oscillators from color Lie (super)algebras}

The graded bracket (\ref{gradedbracket}) introduced in terms of the commutation factors $\varepsilon(\alpha,\beta)$ can be recasted into a linear combination of commutators/anticommutators as 
\bea\label{interpolation}
&\langle A, B\rangle = AB-\varepsilon(\alpha,\beta) BA=  \frac{1}{2}\left(1+\varepsilon(\alpha,\beta) \right)\cdot [A,B] + \frac{1}{2}\left(1-\varepsilon(\alpha,\beta) \right)\cdot \{A, B\}.&
\eea
$\langle A, B\rangle $ is,  for $\varepsilon(\alpha,\beta)\neq \pm 1$, a genuine {\it mixed-bracket} which does not coincide (up to a normalization) with an ordinary $[A, B] $ commutator or $\{A, B\}$ anticommutator. In the following the term {\it mixed-bracket} will be employed for color Lie (super)algebras which present, for at least one choice of $\alpha,\beta$, a commutation factor
$\varepsilon(\alpha,\beta)\neq \pm 1$. The notion of mixed-bracket applies, in particular, to $\varepsilon(\alpha,\beta)$ 
given by roots-of-unity of level-$k$ with $k>2$ (see Appendix {\bf B} for the notion root-of-unity level).\par
In this Section we introduce mixed-bracket para-oscillators from color Lie (super)algebras; they are given by (color)  mixed-bracket generalizations of the ordinary Heisenberg-Lie (super)algebras.  It is important, for the physical applications discussed in the following, to point out the differences between two broad classes of para-oscillators: the parabosonic oscillators versus the
parafermionic ones. The color Lie (super)algebras under consideration in this paper have ${\mathbb Z}_2^p\times {\mathbb Z}_3^q$ gradings for integer values of $p,q$.
We introduce here specific examples of mixed-brackets para-oscillators which, in the following, will be used to illustrate the physical implications of the parastatistics derived from these color Lie (super)algebras.
~\par
Before proceeding with the construction of mixed-bracket para-oscillators we present two important remarks which clarify the framework.\par
{\it Remark 1}: The mixed-bracket parabosonic and parafermionic oscillators are not directly related to the notion of parabosons and parafermions defined, following \cite{{gre},{grme}}, by the trilinear relations. They are defined in terms of a different mathematical structure.\par
{\it Remark 2}: The color Heisenberg-Lie (super)algebras under consideration act as spectrum-generating (super)algebras to determine energy eigenstates and spectrum of quantum models. Contrary to parabosonic oscillators, the creation operators of the parafermionic oscillators are nilpotent. Due to that, the parafermionic oscillators obey a generalization of the Pauli exclusion principle leading, as in the models discussed in the following,
to truncations of the energy spectra. Parafermionic creation operators are associated with a $-1$ diagonal commutation factor (therefore, see also the comment after formula (\ref{epsilonalgebras}), they require a color Lie superalgebra), while parabosonic creation operators are associated with a $+1$ commutation factor.

\subsection{Parabosonic oscillators from color Lie algebras}

We introduce at first four bosonic oscillators $a_I, a_I^\dagger:= (a_I)^\dagger$ with $I=1,2,3,4$, spanning together with the central charge $c$ the bosonic Heisenberg-Lie algebra ${\mathfrak{h}}_{bos}(4)$.
\bea\label{4bososc}
&&{\mathfrak{h}}_{bos}(4)~ : \quad \{a_I,a_I^\dagger,c\}, \quad ~~~ {\textrm{where}}\nonumber\\
&&[a_I,a_J^\dagger ]=\delta_{IJ}\cdot c, \qquad [c,a_I]=[c, a_I^\dagger] =[a_I,a_J]=[a_I^\dagger,a_J^\dagger]= 0 \qquad \forall I,J.
\eea 
With the help of the $\zthreetwo$ graded matrices (\ref{gradedcomm}) $C_{\underline{ij}}$ we can introduce the
parabosonic color Heisenberg-Lie algebra  ${\mathfrak{h}}_{pb}(4)=\{A_I,A_I^\dagger,C\}$ spanned by $9$ generators given by
\bea\label{genhpb4}
&& A_1 = C_{\underline{20}} \cdot a_1,\qquad\qquad A_1^\dagger = C_{\underline{10}}\cdot a_1^\dagger,\nonumber\\
&&A_2 = C_{\underline{22}} \cdot a_2, \qquad\qquad A_2^\dagger = C_{\underline{11}}\cdot a_2^\dagger, \qquad \nonumber\\
&& A_3 = C_{\underline{21}} \cdot a_3,\qquad\qquad A_3^\dagger = C_{\underline{12}}\cdot a_3^\dagger, \qquad\nonumber\\
&&  {{ A_4 = C_{\underline{02}}\cdot  a_4}},\qquad\qquad A_4^\dagger = C_{\underline{01}} \cdot a_4^\dagger, \qquad\nonumber\\
&&C~ = C_{\underline{00}}.
\eea
Their respective $\zthreetwo$ gradings are expressed by the ${\underline{ij}}$ suffix in the right hand side.\\
It follows that  ${\mathfrak{h}}_{pb}(4)$ is a color Lie algebra satisfying the graded brackets:
\bea\label{hpb4brackets}
&&{\mathfrak{h}}_{pb}(4)~ : \quad \{A_I,A_I^\dagger,C\}, \quad ~~~~~ {\textrm{where}}\nonumber\\
&&\langle A_I,A_J^\dagger \rangle =\delta_{IJ}\cdot C, \qquad \langle C,A_I\rangle=\langle C, A_I^\dagger\rangle =\langle A_I,A_J\rangle=\langle A_I^\dagger,A_J^\dagger\rangle= 0 \qquad \forall I,J.
\eea
The graded brackets are recovered from equation (\ref{gradedbracket}) by inserting the $\varepsilon({{\underline{ij}},{\underline {kl}}})$ commutation factors presented in table (\ref{z3z3array}). The special choice $j=j_1 = e^{\frac{2\pi i}{3}}$ gives to ${\mathfrak{h}}_{pb}(4)$ the status of a mixed-bracket color Heisenberg-Lie algebra.\par
One should note that the four pairs of creation/annihilation operators are constructed  in terms of the pairings
\bea
&
{\underline{01}} \leftrightarrow {\underline{02}},  \qquad
{\underline{10}} \leftrightarrow {\underline{20}},  \qquad
{\underline{11}} \leftrightarrow {\underline{22}},  \qquad
{\underline{12}} \leftrightarrow {\underline{21}}&
\eea
which lead to a ${\underline{00}}$-graded sector and whose commutation factors satisfy
\bea
&\varepsilon({{\underline{01}},{\underline {02}}})= \varepsilon({{\underline{10}},{\underline {20}}})=
\varepsilon({{\underline{11}},{\underline {22}}})=\varepsilon({{\underline{12}},{\underline {21}}})=1.
\eea

\subsection{Parafermionic oscillators from color Lie superalgebras}

The simplest example of a color Lie superalgebra which introduces mixed-bracket parafermionic oscillators is obtained from the ${\mathbb Z}_2\times\zthreetwo$ grading whose commutation factors are presented in table (\ref{z2z3z3array}). The presence of nontrivial mixed brackets requires to set $j=j_1 = e^{\frac{2\pi i}{3}}$, while the presence of parafermions is implied by setting $\delta=-1$. This choice of parameters produces the $iv$ case in the (\ref{summary}) list of (super)algebras.\par
We now proceed to introduce the mixed-bracket color Heisenberg-Lie superalgebra 
${\mathfrak{h}}_{pf}(4|4)$. The following ingredients are used in its construction:\par
1) the $\zthreetwo$ graded matrices $C_{\underline{ij}}$ given in (\ref{gradedcomm});\par
2) the $2\times 2$, $\ztwo$-graded matrices $I$ and $Y$, of respective grading $0$ and $1$, given by
{\footnotesize{\bea
I=\left(\begin{array}{cc} 1&0\\0&1\end{array}\right)\in[0],  &\qquad&
Y=\left(\begin{array}{cc} 0&1\\1&0\end{array}\right)\in[1];\nonumber
\eea
}}

3) the four ordinary bosonic oscillators $a_I, a_I^\dagger$ (for $I=1,2,3,4$) and central charge $c$ which have already been introduced in (\ref{4bososc});\par
4) four ordinary fermionic oscillators $f_I, f_I^\dagger$ (for $I=1,2,3,4$); together with the central charge $c$
they close the fermionic Heisenberg-Lie superalgebra ${\mathfrak{h}}_{fer}(4)$ given by
\bea\label{4ferosc}
&&{\mathfrak{h}}_{fer}(4)~ : \quad \{f_I,f_I^\dagger,c\}, \quad ~~~ {\textrm{where}}\nonumber\\
&&\{f_I,f_J^\dagger \}=\delta_{IJ}\cdot c, \qquad [c,f_I]=[c, f_I^\dagger] =\{f_I,f_J\}=\{f_I^\dagger,f_J^\dagger\}= 0 \qquad \forall I,J.
\eea 
The bosonic and the fermionic oscillators mutually commute:
\bea
&[a_I,f_J]=[a_I, f_J^\dagger] =[a_I^\dagger,f_J]=[a_I^\dagger, f_J^\dagger]= 0 \qquad {\textrm{for all $I,J$.}}&
\eea

The mixed-bracket color Heisenberg-Lie superalgebra ${\mathfrak{h}}_{pf}(4|4)$ is spanned by $17$ generators 
($C, A_I, A_I^\dagger, F_I,F_I^\dagger$ for $I=1,2,3,4$). The subalgebra spanned by $C, A_I,A_I^\dagger$ is parabosonic and isomorphic to ${\mathfrak{h}}_{pb}(4)$. The subalgebra spanned by $C, F_I,F_I^\dagger$
defines four pairs of parafermionic oscillators. A realization of the ${\mathfrak{h}}_{pf}(4|4)$ generators is obtained through the following positions:
\bea\label{genhpf4}
&&
C~ = I\otimes C_{\underline{00}},\nonumber\\
&& A_1 = I\otimes C_{\underline{20}} \cdot a_1,\qquad\qquad A_1^\dagger = I\otimes C_{\underline{10}}\cdot a_1^\dagger,\nonumber\\
&&A_2 = I\otimes  C_{\underline{22}} \cdot a_2, \qquad\qquad A_2^\dagger = I\otimes C_{\underline{11}}\cdot a_2^\dagger, \qquad \nonumber\\
&& A_3 = I\otimes C_{\underline{21}} \cdot a_3,\qquad\qquad A_3^\dagger = I\otimes C_{\underline{12}}\cdot a_3^\dagger, \qquad\nonumber\\
&&  A_4 = I\otimes C_{\underline{02}}\cdot  a_1,\qquad\qquad A_4^\dagger = I\otimes C_{\underline{01}} \cdot a_4^\dagger, \qquad\nonumber\\
&& F_1 = Y\otimes C_{\underline{20}} \cdot f_1,\qquad\qquad F_1^\dagger = Y\otimes C_{\underline{10}}\cdot f_1^\dagger,\nonumber\\
&&F_2 = Y\otimes C_{\underline{22}} \cdot f_2, \qquad\qquad F_2^\dagger = Y\otimes C_{\underline{11}}\cdot f_2^\dagger, \qquad \nonumber\\
&& F_3 = Y\otimes C_{\underline{21}} \cdot f_3,\qquad\qquad F_3^\dagger = Y\otimes C_{\underline{12}}\cdot f_3^\dagger, \qquad\nonumber\\
&&  F_4 = Y\otimes C_{\underline{02}}\cdot  f_4,\qquad\qquad F_4^\dagger = Y\otimes C_{\underline{01}} \cdot f_4^\dagger. \qquad
\eea

Since no confusion will arise, for simplicity we employed for the ${\mathfrak{h}}_{pb}(4)$ subalgebra generators the same symbols as those introduced in the set of equations (\ref{genhpb4}).\par
In the $1$ bit -  $2$ trits notation the grading assignment of the (\ref{genhpf4}) generators is:
\bea
&&C\in[0{\underline{00}}], \nonumber\\
&&  A_1^\dagger \in[0{\underline{10}}],~ A_1 \in[0{\underline{20}}], ~A_2^\dagger \in[0{\underline{11}}], ~A_2 \in[0{\underline{22}}], ~A_3^\dagger \in[0{\underline{12}}], ~A_3 \in[0{\underline{21}}], ~A_4^\dagger \in[0{\underline{01}}],~ A_4 \in[0{\underline{02}}], \nonumber\\
&&  F_1^\dagger \in[1{\underline{10}}],~ F_1 \in[1{\underline{20}}], ~F_2^\dagger \in[1{\underline{11}}], ~F_2 \in[1{\underline{22}}], ~F_3^\dagger \in[1{\underline{12}}], ~F_3 \in[1{\underline{21}}], ~F_4^\dagger \in[1{\underline{01}}], ~ F_4 \in[1{\underline{02}}].\nonumber\\
&&
\eea

The color Heisenberg-Lie superalgebra ${\mathfrak{h}}_{pf}(4|4)$ closes the following set of graded brackets which are defined in terms of the commutation factors entering table (\ref{z2z3z3array}). We have
\bea\label{hpf4brackets}
&&{\mathfrak{h}}_{pf}(4|4)~ : \quad \{A_I,A_I^\dagger,F_I,F_I^\dagger, C\}, \quad ~~~~~ {\textrm{where, ~~$\forall I,J,$}}\nonumber\\
&&
\langle C, C\rangle= \langle C,A_I\rangle=\langle C, A_I^\dagger\rangle =\langle C,F_I\rangle=\langle C,F_I^\dagger\rangle= 0,\nonumber\\
&&\langle A_I,A_J^\dagger \rangle =\delta_{IJ}\cdot C, \qquad \langle A_I,A_J\rangle=\langle A_I^\dagger,A_J^\dagger\rangle= 0,\nonumber\\
&&\langle F_I,F_J^\dagger \rangle =\delta_{IJ}\cdot C, \qquad ~\langle F_I,F_J\rangle=\langle F_I^\dagger,F_J^\dagger\rangle= 0,\nonumber\\
&&\langle A_I,F_J^\dagger \rangle =\langle A_I,F_J^\dagger\rangle=\langle A_I^\dagger,F_J\rangle=\langle A_I^\dagger, F_J^\dagger\rangle= 0.
\eea

From formula (\ref{hpf4brackets}) one should note, in particular, the nilpotency of the parafermionic creation operators induced by the $\langle F_I^\dagger,F_I^\dagger\rangle=0$ bracket. Indeed, we get
\bea\label{nilpotency}
(F_I^\dagger)^2&=&0 \qquad{\textrm{~~~ for any $I=1,2,3,4$.}}
\eea

One should note that one of the $18$ graded sectors of ${\mathbb Z}_2\times\zthreetwo$ is left empty
($\emptyset\in [1{\underline{11}}]$), while each one of the remaining $17$ graded sectors accommodate a single generator of ${\mathfrak{h}}_{pf}(4|4)$. It would be tempting to add to $[1{\underline{11}}]$ an extra generator such as ${\overline C}= Y\otimes C_{\underline{00}}$. The introduction of ${\overline C}$ would require to further introduce extra generators in the algebra since, e.g., $\langle {\overline C}, F_1\rangle =\{{\overline C}, F_1\}= 2\cdot I\otimes F_1\in 
[0{\underline{20}}]$ while, on the other hand, $\langle {\overline C}, F_1\rangle \not\propto A_1\in 
[0{\underline{20}}]$. \par
The classification of the {\it minimal}, inequivalent ${\mathbb Z}_2\times{\mathbb Z}_2$-graded  color Lie algebras and color Lie superalgebras was presented in \cite{kuto}. In this context the notion of {\it minimal} refers to the (super)algebras being spanned by one and only one generator in each graded sector. 
The superalgebra ${\mathfrak{h}}_{pf}(4|4)$ is not minimal due to its empty sector $ [1{\underline{11}}]$.  On the other hand, the $9$-generator parabosonic Heisenberg-Lie algebra ${\mathfrak{h}}_{pb}(4)$ previously introduced in (\ref{genhpb4},\ref{hpb4brackets}) is an example of a {\it minimal} $\zthreetwo$-graded color Lie algebra.

\section{Intermezzo: on braidings and multi-particle states}

In this Section we anticipate some relevant features of colored paraoscillators; the aim  is to provide an introduction to the more detailed discussions (presented in the construction of specific models) which will be given in the following.
~\par
The parabosonic $A_I^\dagger$ and the parafermionic  $F_I^\dagger $  creation operators, respectively introduced
 in (\ref{genhpb4}) and (\ref{genhpf4}),  belong to a larger class of non-(anti)commutative creation operators  recovered from color Lie (super)algebras whose graded brackets (\ref{interpolation}) are of mixed type (i.e., admitting commutation factors $\varepsilon(\alpha,\beta)\neq \pm 1$). Let's denote with $D_J^\dagger$ the creation operators belonging to this broader class. We get, in particular, for $J\neq J'$:
\bea
D_{J}^\dagger D_{J'}^\dagger &=& a_{J,J'}\cdot D_{J'}^\dagger D_{J}^\dagger,
\eea
where the nonvanishing $a_{J,J'}$ parameters encode the non (anti)commutative properties of the creation operators. By specializing to three operators ($J,J'=1,2,3$) we can set, e.g.,
\bea\label{threecreation}
&D_1^\dagger D_2^\dagger = a\cdot D_2^\dagger D_1^\dagger,\qquad 
D_2^\dagger D_3^\dagger = b \cdot D_3^\dagger D_2^\dagger,\qquad
D_3^\dagger D_1^\dagger = c\cdot D_1^\dagger D_3^\dagger,&
\eea
where the nonvanishing parameters $a,b,c$ are given by certain commutation factors.
By identifying the $D_J^\dagger$'s with three creation operators from (\ref{genhpb4}) and (\ref{genhpf4}) we respectively have
\bea\label{specialcases}
&&D_J^\dagger \equiv A_J^\dagger\quad {\textrm{~~ for ~~$J=1,2,3$, ~~~ implying that \quad $a=b=c=j$}},
\nonumber\\
&&D_J^\dagger \equiv F_J^\dagger\quad {\textrm{~~ for ~~$J=1,2,3$, ~~~ implying that \quad $a=b=c=-j$}}.
\eea

The three creation operators $D_J^\dagger$ from (\ref{threecreation}) can either be assumed to be parabosonic (hence, 
$(D_J^\dagger)^2\neq 0$) or parafermionic.\par
The parafermionic operators are nilpotent: $(D_J^\dagger)^2=0$ for $J=1,2,3$. Due to the nilpotency, in the parafermionic case the  powers of their symmetric linear combinations vanish for $n\geq 4$, so that 
\bea
(D_1^\dagger+D_2^\dagger+D_3^\dagger)^n&=&0 \qquad{\textrm{for $n\geq 4$.}}
\eea 
At the special $n=3$ value we have, for parafermionic oscillators, 
\bea
(D_1^\dagger+D_2^\dagger+D_3^\dagger)^3&=&w(a,b,c) \cdot D_1^\dagger D_2^\dagger D_3^\dagger, \qquad {\textrm{where}} \nonumber\\
w(a,b,c) &=&1+a^{-1}+b^{-1}+ca^{-1}+cb^{-1}+ca^{-1}b^{-1}.
\eea
The right hand side is nonvanishing apart from special $a,b,c$ values producing the ``miraculous cancellation"
\bea\label{cancellation}
w(a,b,c)&=& 0.
\eea
Let's consider some cases.\par
If $a=b=c=-1$ we have ordinary fermions.
For these values, due to the Pauli exclusion principle, even the 2nd power vanishes:
$(D_1^\dagger+D_2^\dagger+D_3^\dagger)^2 =0$. 
\par
For the parafermionic creation operators defined in (\ref{genhpf4}),  $a=b=c=-j_1=-e^{\frac{2\pi i}{3}}$, there is no truncation at the 3rd power:
$$ w(a,b,c)=3(1-j_1^2)\neq 0,\quad and\quad  (D_1^{\dagger}+D_2^{\dagger}+D_3^{\dagger})^3\neq 0 .$$
\par
A 3rd order truncation (\ref{cancellation}) is obtained from $a,b,c$ 
given by
$a=b=j_1$, $c=j_1^2$. Indeed:
\bea\label{truncation}
w(j_1,j_1,j_1^2)&=&0  \qquad {\textrm{for $ j_1=e^{\frac{2\pi i}{3}}$.}}
\eea

\subsection{Braiding properties}

We illustrate the braiding properties of the mixed-bracket colored paraoscillators by analyzing the example of 
three parabosonic creation operators $A_J^\dagger$ introduced in (\ref{genhpb4}) and whose mutual commutation factors are given in (\ref{specialcases}). We set $j\equiv j_1=e^{\frac{2\pi i}{3}}$.\par
~\par
The $6$-generator ${\bf S}_3$ permutation group acts with generators $S_{12}$, $S_{23}$ defined as
\bea
S_{12}:~  A_1^\dagger \leftrightarrow A_2^\dagger, \quad A_3^\dagger \mapsto A_3^\dagger,  &\qquad&
S_{23}:~  A_1^\dagger \mapsto A_1^\dagger, \quad A_2^\dagger \leftrightarrow A_3^\dagger.  
\eea
The generators  $S_{12}$, $S_{23}$ satisfy
\bea
S_{12}^2=  S_{23}^2={\bf 1},   &\quad&  S_{12}\cdot S_{23}\cdot S_{12}= S_{23}\cdot S_{12}\cdot S_{23}.  
\eea
The action on symmetric polynomials such as $(A_1^\dagger+A_2^\dagger+A_3^\dagger)^n$ obviously produces the identity
\bea
&S_{12}\left((A_1^\dagger+A_2^\dagger+A_3^\dagger)^n\right)=S_{23}\left((A_1^\dagger+A_2^\dagger+A_3^\dagger)^n\right)=(A_1^\dagger+A_2^\dagger+A_3^\dagger)^n.&
\eea
On the other hand, the action on non-symmetric polynomials such as $A_1^\dagger A_2^\dagger$, $ A_1^\dagger A_2^\dagger A_3^\dagger$ induces a braid group representation with $S_{12} \rightarrow B_{12}$, $S_{23}\rightarrow B_{23}$. Indeed, we get
\bea
B_{12}(A_1^\dagger A_2^\dagger) = A_2^\dagger A_1^\dagger = j_1^2\cdot A_1^\dagger A_2^\dagger,\qquad  B_{12}^2(A_1^\dagger A_2^\dagger)= j_1\cdot A_1^\dagger A_2^\dagger, \qquad B_{12}^3(A_1^\dagger A_2^\dagger)=  A_1^\dagger A_2^\dagger&& \quad
\eea
and 
\bea\label{braid3}
&B_{12}(A_1^\dagger A_2^\dagger A_3^\dagger) = 
B_{23}(A_1^\dagger A_2^\dagger A_3^\dagger)= j_1^2\cdot A_1^\dagger A_2^\dagger A_3^\dagger, &
\eea
so that  $B_{12}\equiv B_{23}$ satisfy 
\bea 
B_{12}\cdot B_{23}\cdot B_{12}= B_{23}\cdot B_{12}\cdot B_{23}&~~&
{\textrm{(with  $B_{12}^3= B_{23}^3={\bf 1}$).}}
\eea
Formula (\ref{braid3}) gives a trivial representation of the braid group ${\bf B}_3$.
\par
For {\it mixed-bracket} commutation factors, a case by case analysis to model-construction should be made.  It has to specify whether the paraparticles introduced in a specific model obey a parastatistics based on the permutation group or, alternatively, on the braid group. 

\subsection{On multi-particle parastatistics}

Two different frameworks, which have been applied in the literature to multi-particle parastatistics, can be extended to mixed-bracket color Lie (super)algebras.\par
At first we discuss how to introduce First-Quantized multi-particle sectors from graded Hopf algebras endowed with a braided tensor product \cite{maj}. \par
A color Heisenberg-Lie (super)algebra ${\mathfrak h}$  induces a color Universal Enveloping Algebra $U\equiv {\cal U} ( {\mathfrak{h}})$ which has a status of a graded Hopf algebra endowed with a braided tensor product (see \cite{rbzha} and references therein for a recent account on color Hopf algebras). A natural framework to induce multi-particle quantization from single-particle quantum Hamiltonians consists in constructing multi-particle states
in terms of Hopf algebra (co)structures; a particular role is played by the $\Delta: U\rightarrow U\otimes U$ coproduct. In \cite{{top1},{top2},{nbits}}  this framework was applied to ${\mathbb Z}_2^n$-graded color Lie (super)algebras; it was used to prove the existence of signatures of $n$-bit parastatistics, i.e., of measurements whose results cannot be reproduced from ordinary bosons/fermions.\par
\par
The coassociativity of the coproduct, expressed as
\bea\label{coproducts}
\Delta^{(N+1)} &:=& (\Delta \otimes {\bf 1})\Delta^{(N)}=({\bf 1}\otimes \Delta)\Delta^{(N)}\quad{\textrm{(where $\Delta^{(1)}\equiv \Delta$), \quad implies }}\nonumber\\
\Delta^{(N)} &:& U\rightarrow U^{\otimes^{N+1}}=U\otimes \ldots \otimes U {\textrm{\quad ~~(the tensor product of $N+1$ spaces)}}.
\eea

Starting from a single-particle Hilbert space ${\cal H}^{(1)}$, spanned by the $D_J^\dagger\in {\mathfrak h}$ creation operators applied to the single-particle vacuum $|vac\rangle_{(1)}$, the coproduct 
allows to construct the First-Quantized $N$-particle Hilbert space ${\cal H}^{(N)}$. We get 
\bea {\cal H}^{(N)}\subset {{\cal H}^{(1)}}^{\otimes^N}, && {\textrm{with ${\cal H}^{(N)}$ spanned by the $\Delta^{(N-1)} (D_J^\dagger)$ creation operators applied}} \nonumber\\ &&{\textrm{to the $N$-particle vacuum  $|vac\rangle_{(N)}=|vac\rangle_{(1)}\otimes \ldots \otimes |vac\rangle_{(1)}$.}}
\eea 
The consistency of this approach is implied by the braided tensor product. We illustrate it with the example of the parabosonic creation oscillators $A_1^\dagger, A_2^\dagger$ introduced in (\ref{genhpb4}); they satisfy
$A_1^\dagger A_2^\dagger = j_1 A_2^\dagger A_1^\dagger$ and possess a vanishing graded bracket: $
\langle A_1^\dagger, A_2^\dagger\rangle =0$. The braided tensor product gives
\bea
({\bf 1}\otimes A_1^\dagger)\cdot (A_2^\dagger \otimes {\bf 1})&=& j_1 (A_2^\dagger\otimes {\bf 1})\cdot ({\bf 1}\otimes A_1^\dagger) = j_1 A_2^\dagger \otimes A_1^\dagger,\nonumber\\
({\bf 1}\otimes A_2^\dagger)\cdot (A_1^\dagger \otimes {\bf 1})&=& j_1^2 (A_1^\dagger\otimes {\bf 1})\cdot ({\bf 1}\otimes A_2^\dagger) = j_1^2 A_1^\dagger \otimes A_2^\dagger.
\eea

Straightforward computations guarantee covariant formulas for the $2$-particle creation operators induced by the coproduct. It is easily shown that, introducing the $2$-particle graded bracket $\langle \Delta(A_1^\dagger), \Delta(A_2^\dagger)\rangle_{(2)} $ as
\bea
\langle \Delta(A_1^\dagger), \Delta(A_2^\dagger)\rangle_{(2)}&:=& \Delta(A_1^\dagger)\Delta(A_2^\dagger)-j_1\Delta(A_2^\dagger)\Delta(A_1^\dagger),
\eea
the following relation is implied:
\bea
\langle A_1^\dagger, A_2^\dagger\rangle =0 &\Rightarrow&\langle \Delta(A_1^\dagger), \Delta(A_2^\dagger)\rangle_{(2)} = \Delta (
\langle A_1^\dagger, A_2^\dagger\rangle)  =0.
\eea

The connection between the \cite{maj} graded Hopf algebra approach to parastatistics and the traditional \cite{{gre},{grme}} framework based on trilinear relations has been discussed in \cite{{anpo1},{kada1}}. \par
~\par
A different scheme (already employed in \cite{topvoli}
to obtain the quantization of braided Majorana qubits) is applied to introduce {\it indistinguishable} particles from $N$ creation operators.\par
 We illustrate its main features by taking, as an example, $N=3$ parabosonic creation operators $A_J^\dagger$ ($J=1,2,3$) from (\ref{genhpb4}) and whose noncommutative relations are given in (\ref{specialcases}). \par
The introduction of a Hilbert space ${\cal H}$, spanned by the vectors
\bea
&&(A_1^\dagger)^{n_1} (A_2^\dagger)^{n_2}(A_3^\dagger)^{n_3}  |vac\rangle\in {\cal H} {\textrm{\qquad for $n_1,n_2,n_3=0,1,2,\ldots $,}}\nonumber\\
&&{\textrm{with $|vac\rangle$ being a Fock's vacuum state satisfying $A_J|vac\rangle=0$ for $J=1,2,3$,}} 
\eea
guarantees that $A_1^\dagger, A_2^\dagger, A_3^\dagger$ create excited states of distinct particles of deformed oscillators with respective coordinates $x,y,z$ (this point will be discussed in Section {\bf 7}).
Following {\cite{topvoli}, an indistinguishability requirement can be imposed as a superselection, demanding that a Hilbert space ${\cal H}_{ind}$ of indistinguishable particles is realized as 
\bea\label{symmetrized3}
&{\cal H}_{ind}\subset {\cal H}, {\textrm{\qquad where  $\quad (A_1^\dagger+A_2^\dagger+A_3^\dagger)^n|vac\rangle\in {\cal H}_{ind}\quad $ for $n=0,1,2,\ldots $.}}
 &
\eea
The parabosonic creation operators $A_J^\dagger$ are used as building blocks to construct the symmetrized three-dimensional wave functions. In connection with the $\zthreetwo$ grading  (respectively given for $A_1^\dagger, ~A_2^\dagger,~ A_3^\dagger$ by the graded sectors $ [\underline{10}]$, $[\underline{11}]$, $[\underline{12}]$), the index ${\underline 1}$ of the first ${\mathbb Z}_3$ grading defines the grading of the symmetrized 
$(A_1^\dagger+A_2^\dagger+A_3^\dagger)$ creation operator, while the second ${\mathbb Z}_3$ grading (respectively, ${\underline 0}$,  ${\underline 1}$ or ${\underline 2}$) labels the respective $x,y,z$ coordinates.\\
This framework is easily extended to construct indistinguishable $N$-dimensional wave functions.\par
Sections {\bf 7} and {\bf 8} apply this indistinguishability framework, while  the investigation of multi-particle quantization from mixed-bracket color Hopf algebras discussed before is left for future works.

\section{The minimal $\zthreetwo$-graded parabosonic oscillator model}

We present in this Section the simplest $\zthreetwo$-graded parabosonic oscillator model derived from the mixed-bracket color Heisenberg-Lie algebra  ${\mathfrak{h}}_{pb}(4)$ introduced in 
(\ref{genhpb4}). We introduce at first the single-particle Hilbert space; we then introduce the superselected Hilbert space of {\it indistinguishable} particles according to the scheme outlined in the previous Section.\par
We set the four $a_I, a_I^\dagger$ bosonic oscillators in (\ref{4bososc}) to be expressed as
\bea
a_I= \frac{1}{\sqrt{2}}(x_I+\partial_{x_I}), && 
a_I^\dagger= \frac{1}{\sqrt{2}}(x_I-\partial_{x_I}) {\textrm{\qquad for $~I=1,2,3,4$,}}
\eea
where the $x_I$'s are real space coordinates. Then, the  (\ref{genhpb4}) generators of ${\mathfrak{h}}_{pb}(4)$ are
$9\times 9$ matrix differential operators acting on $9$-component column vectors. Let $v_i$ (for $i=1,2,\ldots, 9)$ denote the column vector with entry $1$ in the $i$-th position and $0$ otherwise. If we assume $v_1$ to be $[{\underline{00}}]$-graded we get, due to the (\ref{gradedcomm}) definition of the $C_{\underline{ij}}$ matrices, that the $\zthreetwo$ grading of the $v_i$'s vectors is given by
\bea
&v_1\in[{\underline{00}}],~ v_2\in[{\underline{02}}],~v_3\in[{\underline{01}}],
~v_4\in[{\underline{20}}],~v_5\in[{\underline{22}}],~v_6\in[{\underline{21}}],
~v_7\in[{\underline{10}}],~v_8\in[{\underline{12}}],~v_9\in[{\underline{11}}].&\nonumber\\&&
\eea
The $[{\underline{00}}]$-graded Hamiltonian $H_{4d;osc}$ of a four-dimensional matrix quantum oscillator is introduced, for $\hbar=m=\omega=1$, as
\bea\label{4dimosc}
H_{4d;osc}&=& \sum_{I=1}^4 A_I^\dagger A_I = \frac{1}{2}\left(-(\partial_{x_1}^2+\partial_{x_2}^2+\partial_{x_3}^2+\partial_{x_4}^2)+ x_1^2+x_2^2+x_3^2+x_4^2-4\right)\cdot {\mathbb I}_9,
\eea
where $A_I^\dagger, A_I$ are the $\zthreetwo$-graded creation/annihilation oscillators from (\ref{genhpb4}) and  ${\mathbb I}_9$ denotes the $9\times 9$ identity matrix.\par
The operators $A_I^\dagger$ create an excited state of energy $E=1$:
\bea
[H_{4d;osc}, A_I^\dagger] &=& A_I^\dagger.
\eea
A Fock Hilbert space ${\cal H}_{4d}$ is introduced in terms of a normalized, $[{\underline{00}}]$-graded vacuum state 
$\psi_{4d;vac}(x_1,x_2,x_3,x_4)\equiv |0,0,0,0\rangle$:
\bea
A_I \psi_{4d;vac}(x_1,x_2,x_3,x_4) &=& 0, {\textrm{\qquad for $I=1,2,3,4$, ~ implying that}}\nonumber\\
\psi_{4d;vac}(x_1,x_2,x_3,x_4)&=&\frac{1}{\pi} e^{-\frac{1}{2}(x_1^2+x_2^2+x_3^2+x_4^2)}\cdot v_1.
\eea
The Fock Hilbert space ${\cal H}_{4d}$ is spanned by the vectors $|n_1,n_2,n_3,n_4\rangle$ (for $n_I$'s non-negative integers)
 of energy $E= n_1+n_2+n_3+n_4$:
\bea
&&|n_1,n_2,n_3,n_4\rangle \in {\cal H}_{4d}, {\textrm {\quad where \quad $|n_1,n_2,n_3,n_4\rangle =(A_1^\dagger)^{n_1}(A_2^\dagger)^{n_2}(A_3^\dagger)^{n_3}(A_4^\dagger)^{n_4}|0,0,0,0\rangle$}}\nonumber\\
&&{\textrm{\qquad\quad and \quad \quad \quad$H_{4d;osc}|n_1,n_2,n_3,n_4\rangle =(n_1+n_2+n_3+n_4)|n_1,n_2,n_3,n_4\rangle$.}}
\eea
{{Two-dimensional and three-dimensional oscillators can also be recovered from ${\mathfrak{h}}_{pb}(4)$ by considering the ${\mathfrak{h}}_{pb}(2)$, ${\mathfrak{h}}_{pb}(3)$ subalgebras (${\mathfrak{h}}_{pb}(2)\subset {\mathfrak{h}}_{pb}(3) \subset {\mathfrak{h}}_{pb}(4)$) where the $I=1,2$ and $I=1,2,3$ restrictions respectively apply. The minimal signature of a $\zthreetwo$ parastatistics is obtained from ${\mathfrak{h}}_{pb}(2)$ whose generators are $\{A_1,A_2,A_1^\dagger, A_2^\dagger, C\}\in {\mathfrak{h}}_{pb}(2)$. For our scopes it is therefore sufficient to introduce the $2$-dimensional oscillator $H_{2d;osc}$ which, for $x\equiv x_1$, $y\equiv x_2$, is given by}}
\bea\label{2dimosc}
H_{2d;osc}&=& \sum_{I=1}^2 A_I^\dagger A_I = \frac{1}{2}\left(-(\partial_{x}^2+\partial_{y}^2)+ x^2+y^2-2\right)\cdot {\mathbb I}_9.
\eea
The normalized $2$-dimensional vacuum state $\psi_{2d;vac}(x,y)\equiv |0,0\rangle$, satisfying 
\bea
&&   A_1\psi_{2d;vac}(x,y)=A_2\psi_{2d;vac}(x,y)=0, \qquad {\textrm{is}}\nonumber\\
&&\qquad \quad\qquad  \psi_{2d;vac}(x,y)=\frac{1}{{\sqrt{\pi}}} e^{-\frac{1}{2}(x^2+y^2)}\cdot v_1.
\eea
The Hilbert space ${\cal H}_{2d}$ is spanned by the orthonormal vectors $|n,m\rangle$:
\bea\label{nmstates}
&&|n,m\rangle \in {\cal H}_{2d}, {\textrm{\qquad where\quad $n,m=0,1,2,\ldots$;}}\nonumber\\
&& {\textrm{$|n,m\rangle= \frac{1}{\sqrt{n!m!}}(A_1^\dagger)^n(A_2^\dagger)^m|0,0\rangle$, \quad with
\quad $\langle n',m'|n,m\rangle =\delta_{nn'}\cdot \delta_{mm'}.$}}
\eea
The $E_{n,m}$ energy eigenvalues are given by
\bea
H_{2d;osc} |n,m\rangle &=& E_{n,m} |n,m\rangle, {\textrm{\qquad where \qquad $E_{n,m}=n+m$.}}
\eea
The $2$-dimensional oscillator ${\cal H}_{2d}$ produces the minimal signature of a $\zthreetwo$-graded parastatistics when introducing {\it indistinguishable} $2$-particle states as outlined in Section {\bf 6}.

\subsection{A minimal signature of $\zthreetwo$-graded parastatistics}

The indistinguishable Hilbert space ${\cal H}_{ind}$ of two-dimensional wave functions is introduced as a subspace of $H_{2d;osc}$:
\bea
&{\cal H}_{ind}\subset H_{2d;osc} {\textrm{\qquad is spanned by the vectors \qquad $(A_1^\dagger+A_2^\dagger)^n|0,0\rangle \in {\cal H}_{in}$.}}
&
\eea

The energy eigenvalues of the energy eigenstates $(A_1^\dagger+A_2^\dagger)^n|0,0\rangle$ coincide with those
of a ordinary harmonic oscillator with vacuum energy set to $0$:
\bea
H_{2d;osc}(A_1^\dagger+A_2^\dagger)^n|0,0\rangle &=& E_n(A_1^\dagger+A_2^\dagger)^n|0,0\rangle,  {\textrm{\quad where ~ $E_n=n$~ for ~$n=0,1,2,\ldots.$}}
\eea
\par
Up to $n\leq 2$ the normalized energy eigenstates $|\psi_n\rangle\propto (A_1^\dagger+A_2^\dagger)^n|0,0\rangle$, given in terms of the $9$-component vectors $v_i$, are
\bea
E_0 =0: \qquad\quad |\psi_0\rangle
&=&\frac{1}{{\sqrt{\pi}}} e^{-\frac{1}{2}(x^2+y^2)}\cdot v_1.\nonumber\\
E_1=1: \qquad\quad |\psi_1\rangle&=&j^2\frac{1}{{\sqrt{\pi}}}e^{-\frac{1}{2}(x^2+y^2)}(x\cdot v_7+y\cdot v_9).\nonumber\\
E_2=2: \qquad\quad |\psi_2\rangle&=&N_j\frac{1}{{\sqrt{\pi}}}e^{-\frac{1}{2}(x^2+y^2)}\left((2x^2-1)\cdot v_4+j^2(2y^2-1)\cdot v_5 +2xy(1+j^2)\cdot v_6\right),\nonumber\\
&& {\textrm{with \quad $N_j = \frac{1}{\sqrt{z+5}}$ \quad for \quad $z=1+j+j^2$.}}
\eea

One should note in the right hand side the presence of the third root of unity $j$ ($j^3=1$). As pointed out in 
Section {\bf 4}, see (\ref{z3z3cor}), two inequivalent cases are recovered from $j\equiv j_3=1$ (ordinary statistics of indistinguishable bosonic particles) and $j\equiv j_1= e^{\frac{2\pi i}{3}}$ (colored parastatistics of indistinguishable
parabosonic particles). The key issue is whether these two cases can be discriminated, producing a detectable signature of the colored parastatistics with respect to the bosonic statistics. \par

For the model under consideration (the $H_{2d;osc}$ Hamiltonian of two indistinguishable particles), the signature
{\it is not given} by the energy spectrum, since the bosonic and the parabosonic energy spectra coincide. On the other hand, a distinct signature is provided by the probability density $p(x,y)$ of finding the particles around the $x,y$ coordinates.\\
~\\
{\bf The minimal signature}: Let's prepare the $2$-particle system in an $E=n$ energy eigenstate. The
respective bosonic $p_{bos;n}(x,y)$ and parabosonic $p_{pb;n}(x,y)$ probability densities are given by
\bea\label{probdensintro}
p_{bos;n}(x,y)&\equiv& p_{n}(x,y)_{|j=j_3} \quad {\textrm{for \quad ($j_3=1$) \quad ~~and}}\nonumber\\
p_{pb;n}(x,y)&\equiv& p_{n}(x,y)_{|j=j_1}\quad {\textrm{for \quad ($j_1= e^{\frac{2\pi i}{3}}$),~ where}}\nonumber\\
p_{j;n}(x,y)&= & |\langle \psi_n|\psi_n\rangle|^2=||\psi_{n}||.
\eea

The probability densities are normalized:
\bea
&\quad \iint_{-\infty}^{+\infty}dxdy \cdot p_{j;n}(x,y) =1.&
\eea

At the $E=0,1$ energy levels no distinct signature of parastatistics is encountered, since
\bea
p_{pb;0}(x,y)&=& p_{bos;0}(x,y) \qquad {\textrm{and}}\nonumber\\
p_{pb;1}(x,y)&=& p_{bos;1}(x,y).
\eea

On the other hand, $E=2$ is the lowest energy level presenting a distinct signature of parastatistics, due to
\bea\label{E2inequality}
p_{pb;2}(x,y)&\neq& p_{bos;2}(x,y).
\eea

The measurable differences of these two $E=2$ probability densities are analyzed in the next Subsection.\par
{\it Remark}: The (\ref{E2inequality}) inequality has a physical significance leading to measurable consequences. By preparing a system of two indistinguishable (para)oscillators in the second excited state (energy level $E_2=2$), with repeated measurements one can determine whether the system under investigation is composed by ordinary bosons or by colored parabosons.
{This is the minimal theoretical signature produced by the colored $\zthreetwo$-graded parastatistics. It can be spotted by measuring the probability density of two indistinguishable (para)oscillators at energy level $E=2$. }

\subsection{Parabosonic versus bosonic probability densities}

The measurable differences of the parabosonic versus bosonic probability densities are respectively due to  the noncommutativity\slash commutativity of the $A_1^\dagger, A_2^\dagger$ creation operators. \\
The two inequivalent 
($j^3=1$, with $j\equiv j_1,j_2,j_3$ for  $j_1=e^{\frac{2\pi i}{3}}, j_2=j_1^2, j_3=1$) cases are:\\~\\
$~$ {\it i}) {\it the bosonic case}: $~ \quad ~~~~~A_2^\dagger A_1^\dagger = A_1^\dagger A_2^\dagger$; $~~~~~~$ for $j_3=1$;\\
{\it ii}) {\it the parabosonic case}: $\quad A_2^\dagger A_1^\dagger =  j_1^2 A_1^\dagger A_2^\dagger$, $~~~~$ for $j_1=e^{\frac{2\pi i}{3}}$ a primitive third root of unity.\par
~\par
The $(A_1^\dagger+A_2^\dagger)^n$ powers for $n=0,1,2,\ldots$, can be expressed by two respective Pascal's triangles, the ordinary one and a noncommutative one. In order to proceed to their construction and spot their difference we introduce the convenient parameter $z\equiv z{|_j}$ through the algebraic relation
\bea
&z_{|_j} := 1+j+j^2 \quad\Leftrightarrow\quad  {\textrm{either $~~~~z_{j_3}=3~~~$ or $~~~ z_{j_1}=z_{j_2}=0$. $\quad$}}&\nonumber\\
&{\textrm{Therefore, $z$ denotes either $z_{bos}=3$ (bosons) or $z_{pb}=0$ (parabosons).}}&
\eea

The $(A_1^\dagger+A_2^\dagger)^n$ powers are given by
\bea
(A_1^\dagger+A_2^\dagger)^n&=&\sum_{k=0}^n V_{n;k}, {\textrm{\quad where\quad $ V_{n;k}= c_{n;k}(A_1^\dagger)^{n-k}(A_2^\dagger)^{k}$ {\textrm{\quad for \quad $k=0,1,\ldots, n$}}.}}
\eea

The $c_{n;k}$ coefficients are recursively determined from $c_{0;0}=1$:
\bea
V_{n+1;k}&=& A_1^\dagger V_{n;k}+ A_2^\dagger V_{n;k-1}, {\textrm{\qquad\quad  implying}}\nonumber\\
c_{n+1;k} &=& c_{n;k} + j^{2(n-k+1)}c_{n;k-1}
\eea
(in the above formula $c_{n;k}=0$ when $k$ is outside the $k=0,\ldots, n$ range).\\
~\par
The following triangle, which encompasses both bosonic and parabosonic cases, is derived; the
$c_{n;k}$ coefficients (presented as entries of the $n^{th}$ row and $k=0,1,\ldots$ row) are given 
in terms of $j,z$. We get:
\bea&
\begin{array}{cccccccc} 
n=0:\quad&1~~~&&&&&& \\
n=1:\quad&1~~~&1~~~&&&&&\\
n=2:\quad&1~~~&z-j~~~&1~~~&&&&\\
n=3:\quad&1~~~&z~~~&z~~~&1~~~&&&\\
n=4:\quad&1~~~&z+1~~~&z(j+1)~~~&z+1~~~&1~~~&&\\
n=5:\quad&1~~~&2z-j~~~&z(2j+1)+1~~~&z(2j+1)+1~~~&2z-j~~~&1~~~&\\
n=6:\quad& 1~~~&2z~~~&z(2z-j)~~~&z^2+z(j+2)+2~~~&z(2z-j)~~~&2z~~~&1~~~\\
{\ldots}:&\ldots&&&&&
\end{array} &
\eea

By specializing to the $j=1, z=3$ bosonic values we recover the ordinary Pascal  triangle:

\bea&
\begin{array}{ccccccc} 
1~&&&&&& \\
1~&1~&&&&&\\
1~&2~&1~&&&&\\
1~&3~&3~&1~&&&\\
1~&4~&6~&4~&1~&&\\
1~&5~&10~&10~&5~&1~&\\
1~&6~&15~&20~&15~&6~&1~\\
{\ldots}&&&&&&
\end{array} &
\eea

The noncommutative parabosonic triangle is recovered from inserting the $j=j_1, ~z=0$ values. This noncommutative version is presented in the following figure.

\newpage

{\bf Figure 1.} Noncommutative Pascal's triangle obtained from the $j_1=e^{\frac{2\pi i}{3}}$  primitive third root of unity. It finds applications to the corresponding parabosonic statistics. The letter ``$X$" denotes  a $X=-j_1$ coefficient. The presence of the $0$ coefficients is due to the relation $z_{|_{j_1}}=1+j_1+j_1^2=0$:

{\centerline{\includegraphics[width=1.0\textwidth]{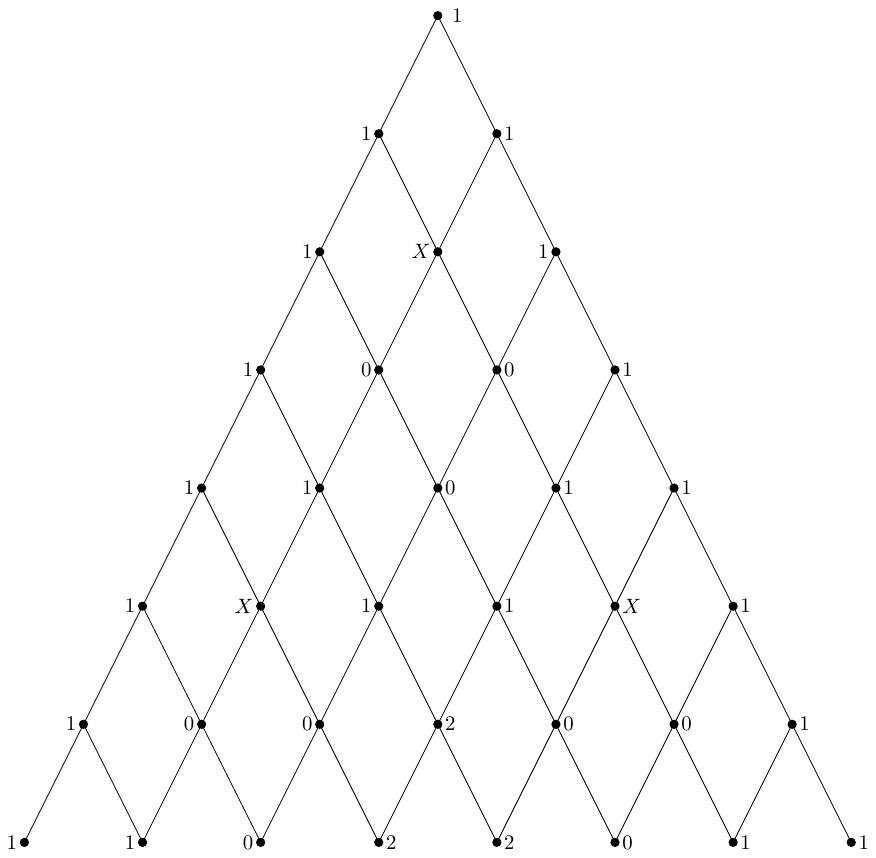}}}

The difference in the commutative versus noncommutative $c_{n;k}$ coefficients entering the respective triangles
has physical implications. It implies that, for $n=2,3,\ldots$, different  normalized $|\psi_n\rangle \propto (A_1^\dagger+A_2^\dagger)^n|0,0\rangle$ symmetrized energy eigenstates are obtained. As discussed in the previous Subsection, this difference is reflected in the measurable probability densities $ p_{j;n}(x,y)$ introduced in (\ref{probdensintro}).\par
The $n=2$ normalized probability density $ p_{j;2}(x,y)$ (associated with the $E_2=2$ energy eigenvalue) can be expressed, in terms of $z$, as
\bea
p_{j;2}(x,y)&=&\frac{1}{(z+5)\pi}e^{-(x^2+y^2)}\left( (2x^2-1)^2+(2y^2-1)^2 +4(z+1)x^2y^2\right).
\eea
The bosonic/parabosonic probability densities $p_{bos;2}(x,y)= p_{j;2}(x,y)_{|_{j=1}}$ and \\$p_{pb;2}(x,y)= p_{j;2}(x,y)_{|_{j=j_1}}$  are respectively recovered from $z=3$ and $z=0$, so that: 
\bea
p_{bos;2}(x,y) &=& \frac{1}{8\pi}e^{-(x^2+y^2)}\left( (2x^2-1)^2+(2y^2-1)^2 +16 x^2y^2\right)\qquad\qquad\quad ~  {\textrm{and}}\nonumber\\
p_{pb;2}(x,y) &=& \frac{1}{5\pi}e^{-(x^2+y^2)}\left( (2x^2-1)^2+(2y^2-1)^2 +4x^2y^2\right).
\eea

~\par
The (\ref{E2inequality}) difference between these two normalized probability densities can be visualized in the following three-dimensional plots, realized with Mathematica.
\newpage

{\bf Figure 2.} Normalized bosonic probability density $p_{bos;2}(x,y)$ in the $-2 \le x,y\le 2 $ range:

~\\

{\centerline{\includegraphics[width=0.7\textwidth]{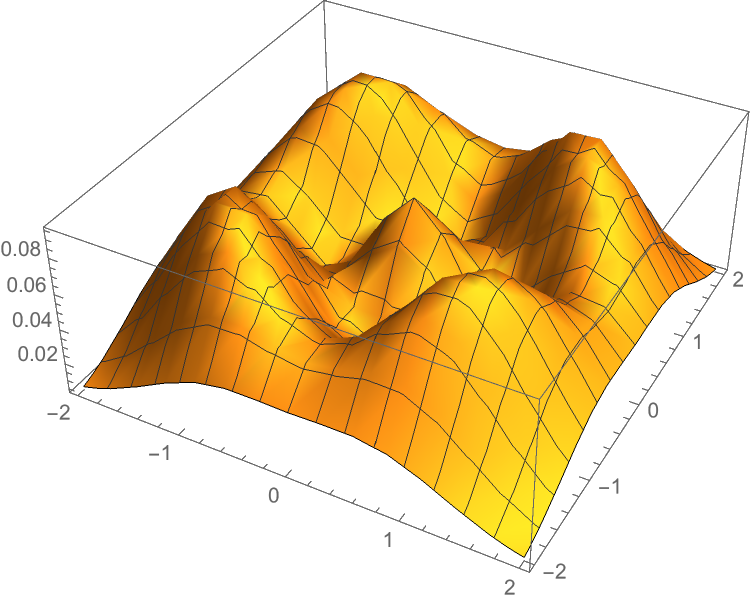}}}
~\\
~\\
~\\
{\bf Figure 3.} Normalized parabosonic probability density $p_{pb;2}(x,y)$ in the $-2 \le x,y\le 2 $ range:

~\\

{\centerline{\includegraphics[width=0.7\textwidth]{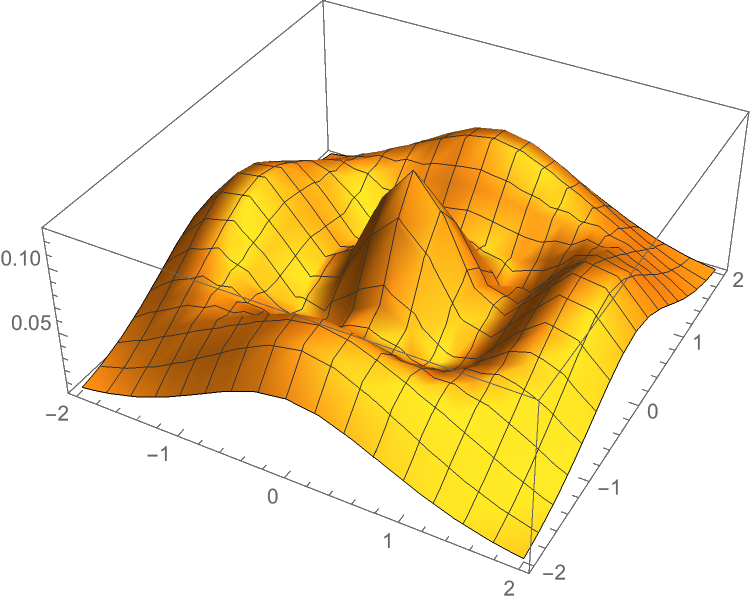}}}

\newpage

One can observe the different shapes of the two probability densities. Both of them admit $5$ local maxima.
We have\\
~\\
{\it i}) in the bosonic case, for $p_{bos}(x,y)$:\\
- one maximum at the origin, given by $p_{bos}(0,0) = \frac{1}{4\pi} \approx 0.079577$ and \\
- $4$ maxima given by $p_{bos}({\overline x},{\overline y})= p_{bos}({\overline x},-{\overline y})= p_{bos}(-{\overline x},{\overline y})= p_{bos}(-{\overline x},-{\overline y})\approx 0.098055,$\\
where ${\overline x}={\overline y}\approx 1.05244 $.\\
~\\
{\it ii}) in the parabosonic case, for $p_{pb}(x,y)$:\\
- one maximum at the origin, given by $p_{pb}(0,0) = \frac{2}{5\pi} \approx 0.127324$ and \\
- $4$ maxima given by $p_{pb}({\overline u},0)= p_{pb}(-{\overline u},0)= p_{pb}(0,{\overline u})= p_{pb}(0,-{\overline u})\approx 0.089194,$\\
where  ${\overline u}\approx 1.53819 $.\\
~\\
In particular it should be noted that, contrary to $p_{bos}(x,y)$, the parabosonic probability density $p_{pb}(x,y)$ has an absolute maximum  at the origin. The pair of two indistinguishable parabosons tends to be more concentrated in the origin than the two indistinguishable bosons. This feature has measurable consequences. By preparing a system of two indistinguishable (para)oscillators in the second excited state (of energy level $E_2=2$), by repeated measurements one can determine whether the system is composed by ordinary bosons or by parabosons at the primitive third root of unity.

\section{Colored parafermionic oscillators and braided Majorana qubits}

In this Section we present applications of colored parafermionic oscillators to quantum models of indistinguishable particles. We prove that, unlike parabosonic oscillators, parafermions induce truncations of the energy spectrum. 
We illustrate this features introducing two quantum models defined for $N=3$ particles (the generalization to any number $N$ of particles is straightforward). The first quantum model is obtained from a ${\mathbb Z}_2\times \zthree^2$-grading and the second one from a ${\mathbb Z}_2^3\times \zthree^2$-grading. Their different physical implications will be pinpointed.\par
~\par
The first quantum model is ${\mathbb Z}_2\times \zthree^2$-graded; it is defined for a  $3$-oscillator subalgebra of the the ${\mathfrak{h}}_{fer}(4)$
 introduced in (\ref{4ferosc}).
It is realized by the $8\times 8$ matrices defined as
\bea\label{3feroscmat}
&f_1^\dagger =\gamma\otimes I\otimes I, \quad f_2^\dagger= X\otimes\gamma\otimes I,\quad f_3^\dagger = X\otimes X\otimes \gamma,&\nonumber\\
&f_1=\beta\otimes I\otimes I, \quad f_2= X\otimes\beta\otimes I,\quad f_3 = X\otimes X\otimes \beta,&\nonumber\\
&c=I\otimes I\otimes I,~&
\eea
where $I, X, \beta,\gamma$ are $2\times 2$ matrices given  by
{\footnotesize{\bea
&I=\left(\begin{array}{cc} 1&0\\0&1\end{array}\right), \qquad X=\left(\begin{array}{cc} 1&0\\0&-1\end{array}\right), \qquad \beta=\left(\begin{array}{cc} 0&1\\0&0\end{array}\right), \qquad \gamma=\left(
\begin{array}{cc} 0&0\\1&0\end{array}\right).&
\eea
}}
One should note that the  (\ref{4ferosc}) (anti)commutators are recovered since $\{X,\beta\}=\{X,\gamma\}=0$.\\
The (\ref{3feroscmat}) matrices, together with the (\ref{genhpf4}) positions for ${\mathfrak{h}}_{pf}(4|4)$, realize the ${\mathfrak{h}}_{pf}(0|3)\subset{\mathfrak{h}}_{pf}(4|4)$ subalgebra as $72\times 72$ matrices given by the following assignments:
\bea\label{hpf3subalg}
&&
~C = ~ c\otimes C_{\underline{00}},\nonumber\\
&& F_1 = f_1\otimes C_{\underline{20}},\qquad\qquad F_1^\dagger = f_1^\dagger\otimes C_{\underline{10}},\nonumber\\
&&F_2 = f_2\otimes C_{\underline{22}}, \qquad\qquad F_2^\dagger = f_2^\dagger\otimes C_{\underline{11}}, \qquad \nonumber\\
&& F_3 = f_3\otimes C_{\underline{02}},\qquad\qquad F_3^\dagger = f_3^\dagger\otimes C_{\underline{01}},\qquad
\eea
where the $\zthree^2$-graded  matrices $C_{\underline{ij}}$ have been introduced in (\ref{gradedcomm}).\par
For $j\equiv j_1=e^{\frac{2\pi i}{3}}$, the (\ref{hpf3subalg}) matrices realize a colored parafermionic superalgebra where, in particular, the non-anticommutative properties of the three nilpotent creation operators $F_I^\dagger$ are  given by
\bea\label{3creations6}
&(F_I^\dagger)^2=0\qquad\qquad {\textrm{for \quad $I=1,2,3$ \quad and\qquad \qquad\qquad}}&\nonumber\\
&F_1^\dagger F_2^\dagger = -j_1 \cdot F_2^\dagger F_1^\dagger, \qquad F_2^\dagger F_3^\dagger = -j_1 \cdot F_3^\dagger F_2^\dagger,\qquad F_1^\dagger F_3^\dagger = -j_1 \cdot F_3^\dagger F_1^\dagger.&
\eea
One should note, in the second line, the common  $(-j_1)$ normalization factor entering the right hand sides. This common factor is a {\it primitive } 6th root of unity satisfying $(-j_1)^3=1$; we recall, following the Appendix {\bf B} conventions, that the  $6$ different 6th roots of unity are split into:
the level-$1$ root $1$, the level-$2$ root $-1$,  two level-$3$ roots given by $j_1, j_1^2$ and, finally, the two primitive level-$6$ roots given by $-j_1, -j_1^2$.\par
~\par
The second quantum model is ${\mathbb Z}_2^3\times \zthree^2$-graded; it is introduced as follows.
At first one replaces the set of three fermionic oscillators from (\ref{3feroscmat}) with a set of three ${\mathbb Z}_2^3$-graded parafermionic oscillators defined by the following $8\times 8$ matrices:
\bea\label{3paraferoscmat}
&p_1^\dagger =\gamma\otimes I\otimes I, \quad p_2^\dagger= I\otimes\gamma\otimes I,\quad p_3^\dagger = I\otimes I\otimes \gamma,&\nonumber\\
&p_1=\beta\otimes I\otimes I, \quad p_2= I\otimes\beta\otimes I,\quad p_3 = I\otimes I\otimes \beta,&\nonumber\\
&c=I\otimes I\otimes I.~&
\eea
An example of their ${\mathbb Z}_2^3$-grading is recovered from table $3_5$ presented in \cite{nbits}, where
$$ p_1,p_1^\dagger\in [100], \quad p_2,p_2^\dagger \in [010], \quad p_3,p_3^\dagger \in [001], \quad c\in [000].$$
The ${\mathbb Z}_2^3$-graded Lie superalgebra ${{\mathfrak{h}}}_{pf}^\ast(3):=\{p_1,p_1^\dagger,p_2,p_2^\dagger,p_3,p_3^\dagger,c\}$ is defined by the \\(anti)commutators:
\bea
&\{p_i,p_i\}=\{p_i^\dagger,p_i^\dagger\}=0 {\textrm{\quad and \quad $\{p_i,p_i^\dagger\} =c$\qquad for any\quad $i=1,2,3$,}}&\nonumber\\
&\qquad [p_i,p_j]=[p_i,p_i^\dagger]=[p_i^\dagger,p_j^\dagger]=0  {\textrm{\qquad \quad~~  for any \quad $i\neq j$,}}&\nonumber\\
&\qquad\qquad\qquad\qquad\qquad   [c,w ] =0 {\textrm{\qquad \qquad\qquad\quad~ for any \quad $w\in {\mathfrak{h}}_{pf}^\ast(3).$ }}&
\eea
Next, the  ${\mathbb Z}_2^3\times \zthree^2$-graded parafermionic superalgebra ${\overline {{\mathfrak{h}}}}_{pf}(3):=\{P_1,P_1^\dagger,P_2,P_2^\dagger,P_3,P_3^\dagger,C\}$ is introduced in terms of the $72\times 72$ matrices 
\bea\label{hpf3s3}
&&
~C = ~ c\otimes C_{\underline{00}},\nonumber\\
&& P_1 = p_1\otimes C_{\underline{20}},\qquad\qquad P_1^\dagger = p_1^\dagger\otimes C_{\underline{10}},\nonumber\\
&&P_2 = p_2\otimes C_{\underline{22}}, \qquad\qquad P_2^\dagger = p_2^\dagger\otimes C_{\underline{11}}, \qquad \nonumber\\
&& P_3 = p_3\otimes C_{\underline{02}},\qquad\qquad P_3^\dagger = p_3^\dagger\otimes C_{\underline{01}}.\qquad
\eea
In the $3$ bits - $2$ trits notation the grading of the  ${\overline{{\mathfrak{h}}}}_{pf}(3)$ generators is assigned to be {{\footnotesize{\bea
&C\in [000\underline{00}], ~ P_1\in [100\underline{20}], ~ P_1^\dagger = ~[100\underline{10}],  ~P_2= [010\underline{22}], ~P_2^\dagger = [010\underline{11}], ~ P_3\in [001\underline{02}], ~P_3^\dagger = [001\underline{01}].&
\eea}}}
The non-anticommutative properties of the three nilpotent creation operators $P_I^\dagger$ are  given by
\bea\label{3creations3}
&(P_I^\dagger)^2=0\qquad\qquad {\textrm{for \quad $I=1,2,3$ \quad and\qquad \qquad\qquad}}&\nonumber\\
&P_1^\dagger P_2^\dagger = j_1 \cdot P_2^\dagger P_1^\dagger, \qquad P_2^\dagger P_3^\dagger = j_1 \cdot P_3^\dagger P_2^\dagger,\qquad P_3^\dagger P_1^\dagger = j_1 \cdot P_1^\dagger P_3^\dagger.&
\eea
In the second line the common  normalization factor entering the right hand sides is $j_1$, that is a {\it primitive} 3rd root of unity. 
The difference, {\it primitive} 6th root of unity from (\ref{3creations6}) versus {\it primitive} 3rd root of unity from (\ref{3creations3}), implies different truncations of the derived energy spectra.

\subsection{The colored parafermionic quantum models}

We introduce now the respective parafermionic quantum models. \par
~\par
From (\ref{hpf3subalg})  a three-particle oscillator Hamiltonian $H_{3,osc}$ is defined to be
\bea
H_{3,osc} &:=& F_1^\dagger F_1+F_2^\dagger F_2+F_3^\dagger F_3.
\eea

 The Hilbert space ${\cal H}_{3}$ is spanned by the vectors
\bea\label{inds6}
&(F_1^\dagger)^{n_1}(F_2^\dagger)^{n_2}(F_3^\dagger)^{n_3}|vac\rangle_{3}\in {\cal H}_{3} {\textrm{\qquad for $n_1,n_2,n_3 =0,1$.}}&
\eea

The Fock vacuum $|vac\rangle_{3}$ satisfies
\bea
&F_I|vac\rangle_{3}=0 {\textrm{\qquad for any $I=1,2,3$.}}&
\eea

The vacuum is given by $|vac\rangle_{3}=v_1$, the $72$-component vector with entry $1$ in the first position and $0$ otherwise.

The energy eigenvalues  of the model are given by $E=0,1,2,3$; their respective degeneracies are $(1,3,3,1)$. One gets a total number of $1+3+3+1=8$ energy eigenvectors which span the ${\cal H}_{3}$ Hilbert space.\par
We can now apply the superselection outlined in Section {\bf 6} to introduce the Hilbert space ${\cal H}_{3;ind}$ of three indistinguishable particles. We get
\bea \label{inds}
&{\cal H}_{3;ind}\subset {\cal H}_{3}\quad {\textrm{is spanned by \quad $(F_1^\dagger +F_2^\dagger +F_3^\dagger)^{n}|vac\rangle_{3}$ \quad for\quad $n=0,1,2,3$.}}&
\eea It follows that
\bea
&&{\textrm{{The indistinguishable Hilbert space  ${\cal H}_{3;ind}$ is spanned by $4$ energy eigenvectors.}}}\nonumber\\
&&{\textrm{{The energy eigenvalues, given by  $E=0,1,2,3$, are nondegenerate.}}}
\eea
~\par

We proceed in the same way from (\ref{hpf3s3}). A three-particle oscillator Hamiltonian ${\overline H}_{3,osc}$ is introduced as
\bea
H_{3,osc} &:=& P_1^\dagger P_1+P_2^\dagger P_2+P_3^\dagger P_3.
\eea
 The Hilbert space ${\overline {\cal H}}_{3}$ is spanned by the vectors
\bea
&(P_1^\dagger)^{n_1}(P_2^\dagger)^{n_2}(P_3^\dagger)^{n3}|vac\rangle_{3}\in {\overline{\cal H}}_{3} {\textrm{\qquad for $n_1,n_2,n_3 =0,1$,}}&
\eea
with the Fock vacuum $|vac\rangle_{3}\equiv v_1$ satisfying
\bea
&P_I|vac\rangle_{3}=0 {\textrm{\qquad for any $I=1,2,3$.}}&
\eea
As in the previous case the energy eigenvalues  of the model are given by $E=0,1,2,3$, with respective degeneracies  $(1,3,3,1)$, for a total number of $1+3+3+1=8$ energy eigenvectors spanning the ${\overline{\cal H}}_{3}$ Hilbert space.\par
The superselection outlined in Section {\bf 6} introduces the Hilbert space ${\overline{\cal H}}_{3;ind}$ of three indistinguishable particles. We get
\bea \label{inds3}
&{\overline {\cal H}}_{3;ind}\subset {\overline {\cal H}}_{3},\quad {\textrm{is spanned by \quad $(P_1^\dagger +P_2^\dagger +P_3^\dagger)^{n}|vac\rangle_{3}$ \quad for\quad $n=0,1,2,3$.}}&
\eea
It follows that
\bea
&&{\textrm{{The indistinguishable Hilbert space  ${\overline {\cal H}}_{3;ind}$ is spanned by {\underline{only}} $3$ energy eigenvectors}}}\nonumber\\
&&{\textrm{{whose energy eigenvalues, given by  $E=0,1,2$, are nondegenerate.}}}
\eea
The reason of the discrepancy with the previous case (i.e., the truncation of the energy spectrum at $E=3$) is due to the miraculous cancellation (\ref{cancellation}) obtained, see (\ref{truncation}), for $w(j_1,j_1,j_1^2)=0.$ It is traced to the 3rd primitive root of unity entering (\ref{3creations3}) with respect to the 6th primitive root of unity entering (\ref{3creations6}).

\subsection{Reconstructing the braided Majorana qubits}

The colored parafermionic quantum models (\ref{inds}) and (\ref{inds3}) reproduce results of the braided 
Majorana qubits introduced in \cite{{topqubits},{topvoli}}. The main features of braided Majorana qubits are briefly summarized in the Appendix {\bf C}. One should note that (\ref{3creations6}) and (\ref{3creations3}) are special cases (for $N=3$ and, respectively, $s_6, ~ s_3$ being 6th and 3rd  primitive roots of unity) of the general formula (\ref{generalnilpotent}) satisfied by braided Majorana qubits creation operators. This general formula leads to the
(\ref{combinatorics}) combinatorics which is responsible for the (\ref{truncatedenergy}) truncations of the energy spectra of the braided Majorana qubits.\par
The connection is deeper than that; the \cite{topvoli} $2$-particle round brackets satisfied by braided Majorana qubits and reproduced in formula (\ref{roundbracket}) in Appendix {\bf C} are recovered from the Rittenberg-Wyler's (\ref{gradedbracket}) parafermionic colored mixed brackets $\langle \cdot,\cdot\rangle$. The identification goes as follows.\par
The connection of the (\ref{roundbracket}) round brackets 
\bea
\left(X,Y\right)_{\theta_{x,y}}&:= & i \sin (\theta_{x,y}) [X,Y]+ \cos (\theta_{x,y})\{X,Y\}~= ~e^{i\theta_{x,y}}\left(XY+e^{-2i\theta_{x,y}}YX\right)\nonumber
\eea
with the color Lie (super)algebra brackets (\ref{gradedbracket}),
\bea
\langle X, Y\rangle  &=& XY-\varepsilon(x,y)YX,
\eea 
requires the identification 
\bea\label{epsilonidentification}
\langle X, Y\rangle &=& e^{-i\theta_{x,y}} (X,Y)\qquad {\textrm{for\quad $\varepsilon(x,y) = - e^{-2i\theta_{x,y}}$,}}
\eea
which relates $\theta_{x,y}$ angles entering (\ref{roundbracket}) and commutant factors.\par
In particular, the Volichenko-type round bracket algebra (\ref{slevelmixed}) is reconstructed 
from the $\langle \cdot,\cdot \rangle$ mixed brackets of the five generators $C, P_1, P_2, P_1^\dagger, P_2^\dagger$ introduced in  (\ref{hpf3s3}), via the identification:
\bea
C\equiv G_0, ~~ P_1^\dagger \equiv G_{+2}, ~~ P_1\equiv G_{-2}, ~~ P_2^\dagger \equiv G_{+1}, ~~  P_2\equiv G_{-1}.
\eea 
At $k=3$, the $\vartheta_{k=3}$ angle entering (\ref{slevelmixed}) corresponds to the primitive 3rd root of unity commutant factor $\varepsilon(\vartheta_{k=3})=- e^{-2i\frac{5}{6}\pi}=e^{\frac{4}{3}\pi i}=j_2=j_1^2$. \par
Similarly, at $k=6$, the (\ref{slevelmixed}) brackets are recovered from the $\langle \cdot,\cdot \rangle$ mixed brackets of the five generators $C, F_1, F_2, F_1^\dagger, F_2^\dagger$ introduced in  (\ref{hpf3subalg}); the identification is
\bea
C\equiv G_0, ~~ F_1^\dagger \equiv G_{+1}, ~~ F_1\equiv G_{-1}, ~~ F_2^\dagger \equiv G_{+2}, ~~  F_2\equiv G_{-2},
\eea 
where the $\vartheta_{k=6}$ angle corresponds to a sixth primitive root of unity commutant factor $\varepsilon(\vartheta_{k=6})=- e^{\frac{2}{3}i\pi}=-j_1$.\par
The obtained result is both a restriction and a generalization of the multi-particle braided quantization of Majorana qubits:\par
~{\it i}) It is a restriction because it was produced for the $k=3,6$ roots of unity levels and not the
more general values $k=3,4,5,6,7,\ldots $. This restriction is due to the fact that we worked with the grading abelian groups $\ztwo^p\times\zthree^q$. A natural expectation is that the construction of color Lie (super)algebras by tensoring more general multiplicative groups (${\mathbb Z}_m$ for arbitrary integer values $m$) allows to recover the general case of level-$k$ truncations. The analysis of this general case is left for a future work;\par
{\it ii}) It is a generalization because (for $k=3,6$) it allows to construct braided multi-particle quantizations of both 
parabosonic {\it and} parafermionic oscillators as seen, e.g., in formula (\ref{hpf4brackets}). Braided Majorana qubits, on the other hand, only involve parafermionic oscillators.\par
As a final note we mention that the  {\it metaabelianess} condition of {\it Volichenko algebras} introduced in \cite{{lese1},{lese2}} (any three generators $X,Y,Z$ satisfy the trilinear relation $[X,[Y,Z]]=0$ for {\it ordinary} commutators) is replaced, for mixed-bracket color Heisenberg-Lie (super)algebras, by a ``graded version of the metaabelianess condition" such that, for any three generators $X,Y,Z$, the relation $\langle X, \langle Y, Z\rangle\rangle =0 $ is satisfied in terms of the graded brackets $\langle \cdot,\cdot\rangle$. This implies that these color Lie (super)algebras are nilpotent of order $3$.

\section{Conclusions}

In this paper we introduced {\it mixed-bracket} color Heisenberg-Lie (super)algebras based on the $\zthree^2$,
$\ztwo\times\zthree^2$ and $\ztwo^3\times\zthree^2$ abelian groups gradings. They define parabosonic and parafermionic oscillators. \par
For parabosons the creation operators are noncommutative; this implies a noncommutative version of the Pascal's triangle, see {\bf Figure 1}. The physical consequence appears in the probability densities of some multi-particle energy eigenstates (see the comparison of bosonic statistics versus parabosons in the $2$-particle sector at the second excited energy level, respectively presented in {\bf Figures 2} and {\bf 3}).\par
For parafermions the nilpotency of their creation operators implies a generalized Pauli exclusion principle. It leads to truncations of their multi-particle energy spectra which implement a Gentile-type \cite{gen} parastatatistics with at most $k-1$ excited states. We have $k=3$ for $\ztwo^3\times\zthree^2$ and
$k=6$ for $\ztwo\times\zthree^2$.
\par
In Section {\bf 6} we discussed the relevance of {\it mixed-bracket} color Lie (super)algebras to parastatistics, pointing out that these colored structures can accommodate two different types of parastatistics: beyond bosons/fermions in any space dimension which are exchanged under the permutation group, as well as anyonic paraparticles exchanged under the braid group. {{Even if the argument given in Subsection {\bf 6.1} was only made for the abelian reps of the braid group, its validity is more general; this is testified by the connection of mixed-bracket parafermions with braided Majorana qubits whose braiding properties are defined, see (\ref{btmatrix},\ref{braidedrel}), by a nonabelian rep of the braid group.}}\par
~\par
The results presented in this paper suggest deep connections  between color Lie (super)algebras and other mathematical structures:\par
{\it i})  In \cite{topvoli} the truncations of the multi-particle spectra of the braided Majorana qubits have been linked to
representations at roots of unity of the quantum superalgebra  ${\cal U}_q({\mathfrak{osp}}(1|2))$. The special properties of the representations of quantum groups at roots of unity, discussed in \cite{{lus},{dck}}, are well known; the specific root-of-unity reps of ${\cal U}_q({\mathfrak{osp}}(1|2))$ are investigated in \cite{{rbz1},{rbzlee}}. It is worth investigating a direct relation between quantum group reps at roots of unity and mixed-bracket color Lie (super)algebras. \par
{\it ii}) The color Lie (super)algebras based on the abelian groups  $\zthree^2$ and $\ztwo^p\times\zthree^2$ (for $p=1,2,3$) do not require, as mentioned in Appendix {\bf A},  the introduction of the cubic, ternary structures discussed in \cite{{kercqg},{bbk},{gora},{cara},{ker},{akl}}.  On the other hand, once their associated  colored quantum models are introduced, the possibility of ternary structures appearing as {\it hidden symmetries} should be duly investigated. \par
{\it iii}) In his influential 1979 paper \cite{sch} Scheunert proved an important theorem which states that, for any finitely generated abelian group, the  associated color Lie (super)algebra can be mapped, via  a multiplier, to an ordinary Lie 
(super)algebra (sometimes this map is referred to as ``decoloring"). This result generated for many years a negative attitude towards color Lie (super)algebras, dismissing their significance with respect to ordinary Lie (super)algebras. This negative attitude is not justified in the light of several recent results (some of them recalled in this paper) concerning the role of colored graded symmetries. In particular, we discussed here at length the relevance of the colored parastatistics with respect to the ordinary bosons/fermions statistics  in multi-particle sectors. A proper interpretation of the seemingly negative result by Scheunert requires a careful analysis.  For instance, investigating its implications beyond color Lie (super)algebras to the realm of color Hopf algebras and their costructures.\par
~\\

{\bf{\Large{Appendix A: a state-of-the-art account of related topics}}}
~\par~\par

To put our paper in due context we present a state-of-the-art account of mathematical and physical applications related with color Lie (super)algebras and their induced parastatistics.\par~
\par
 $1$ {\it - On color Lie (super)algebras}\par
At the beginning, it was mostly the mathematical properties of color Lie (super)algebras which started being investigated, see e.g. \cite{{grja},{cso},{pon}}. Their parastatistics was analyzed in a series of papers \cite{{yaji1},{yaji2},{kaha},{kan},{tol2},{stvdj}}. Despite of that, the applications of color Lie (super)algebras to physics were hampered by a widely widespread (wrong) assumption that their parastatatistics could always be recovered by ordinary bosons/fermions. This situation radically changed in the last decade, with several works investigating different aspects  and applications of ${\mathbb Z}_2^n$-graded
color Lie (super)algebras.  It was recognized \cite{{aktt1},{aktt2}} that they appear as symmetries of known physical systems describing nonrelativistic spinors; a ${\mathbb Z}_2^2$-graded invariant quantum Hamiltonian which prompted further investigations was proposed \cite{brdu};  classical \cite{{akt1},{bru}} and quantum \cite{akt2} systems started being systematically analyzed;  superspace formulation \cite{aikt} and integrable systems \cite{{bru2},{aiktt}} were presented and so on (the list is growing). It was finally proved in \cite{top1} that ${\mathbb Z}_2^2$-graded parafermions  are theoretically detectable, presenting a distinct signature which discriminates them from ordinary fermions; soon after, the detectability of ${\mathbb Z}_2^2$-graded parabosons was established in  \cite{top2}. More works on ${\mathbb Z}_2^n$-graded parastatistics followed \cite{{stvdj2},{nbits},{stvdj3}}.\par
This recent activity on physical applications focused on ${\mathbb Z}_2^n$-graded color Lie (super)algebras, leaving aside the most general class of gradings which induce {\it mixed brackets}. Concerning the mathematical structure, a recent paper \cite{rbzha} presents the updated review of mathematical properties of color Lie (super)algebras and Hopf algebras graded by arbitrary abelian groups.  \\
~\par
$2$ {\it About permutation-group parastatistics} ({\it beyond bosons/fermions in any space dimension})\par
It was commonly believed, see the \cite{conventionality} {\it conventionality of parastatistics argument} based on a localization principle such as \cite{doro}, that paraparticles exchanged under the permutation group could not be experimentally detected (basically,  under a localization hypothesis, all their measurements can be reproduced by ordinary bosons/fermions). It was recently proved that this is not the case. The first test producing a signature for the theoretical detectability of paraparticles exchanged under the permutation group was presented in \cite{top1}; further tests were presented, in chronological order, in \cite{top2}, \cite{waha} and \cite{nbits}. In all these papers the localization hypothesis is evaded (with different mechanisms at work).  As recalled in \cite{ijgmmp}, the experimental detectability of paraparticles is still an open challenge; the experimentalists have methods to engineer paraparticles in the laboratory, see e.g., \cite{{parasim},{paraexp}}. What is lacking, so far, is to put to experimental test the models (such as those presented in \cite{{top1},{top2},{waha},{nbits}}) which present a clear theoretical signature of parastatistics.
About connections with color Lie (super)algebras: the  \cite{{top1},{top2},{nbits}} theoretical tests are for ${\mathbb Z}_2^n$-graded parastatistics.  In \cite{waha} four classes of theories, with $R$-matrices squaring to the Identity, are investigated. The first class corresponds, see  the \cite{waha} Supplementary Informations, to the $\ztwo^2$-graded parafermions investigated in \cite{top1}.  The three remaining classes and their related $R$-matrices do not have a clear mathematical interpretation.  It is pointed out  in \cite{rbzha} that $R$-matrices squaring to the Identity
are recovered from color Lie (super)algebras graded by general abelian groups. This opens the possibility that the
three extra classes in \cite{waha} could be related to mixed-bracket color Lie (super)algebras. It is a line of investigation which deserves being pursued.\par
In a recent parallel development, a permutation-group parastatistics defined by a primitive third root of unity is discussed in \cite{iqoqi}.\par
Concerning applications, recently the possibility of using $\ztwo^2$-graded parafermions for quantum computations has been advocated in \cite{qcz2z2}. \par
~\par

$3$ {\it - On anyons and Topological Quantum Computation}\par

 The notion of {\it $\ztwo$-graded Majorana qubits} was introduced in \cite{topqubits} (see also \cite{topvoli}) and their braiding properties analyzed. In the $\ztwo$-graded qubit the excited state is fermionic and coincides with its own antiparticle (hence, it is a Majorana fermion).\par
The main idea behind \cite{{topqubits},{topvoli}} was to make contact with the Kitaev's program \cite{kit} of Topological Quantum Computation implemented by emergent braided Majorana particles; it offers topological protection from the quantum decoherence of ``ordinary" quantum computers.  The Kitaev's program was also discussed in \cite{{nssfds},{brki}}; the ``knot logic" behind topological quantum computation with Majorana fermions was presented in \cite{kau}.  As advocated in \cite{minimaltqc}, the manipulation of Majorana qubits could offer a minimal setting to implement topological quantum computers, paralleling the manipulation of bits for ordinary computers and of qubits for the present-day quantum computers. An exciting advance is the recent Microsoft's
announcement  \cite{microsoft} of the first quantum chip powered by a topological architecture. It leads, see the \cite{roadmap} roadmap to fault tolerant quantum computation, to a practical implementation of the Kitaev's proposal.
The possibility to use ${\mathbb Z}_3$-symmetric spin chains of one-dimensional parafermions for decoherence-free quantum computation was already advocated in \cite{alfe}.

The possibility in low space dimensions of a parastatistics based on the braid group was first pointed out in \cite{lemy}. These paraparticles were  named {\it anyons} in \cite{wil}. For a historical account of their discovery and the relevant bibliography one can consult \cite{gol}.\par
Unlike the paraparticles exchanging under the permutation groups, anyons have been experimentally detected. The first experimental evidence was presented in 
\cite{expanyons}; the first evidence of non-abelian anyons transforming under higher dimensional representations 
of the braid group was given in \cite{nonabanyons}. One-dimensional anyons have also been experimentally detected,  see \cite{1danyons}. \par
~\par
$4$ {\it - About ${\mathbb Z}_3$-grading and ternary structures}\par
It is pointed out  in Section {\bf 2} that no nontrivial color Lie algebra is obtained from the ${\mathbb Z}_3$ grading group. In the literature the ${\mathbb Z}_3$  grading group has been applied to the so-called ``ternary structures" producing, e.g., the cubic root of the Dirac equation \cite{kercqg}. In ternary mathematics, besides quadratic multiplications, cubic relations are imposed
(see, for instance, the papers \cite{{bbk},{gora}} and the references therein); their physical applications are discussed, e.g., in \cite{{cara},{ker}}. This ternary, ${\mathbb Z}_3$-graded, mathematics/physics is not directly related to color Lie (super)algebras. Nevertheless, $\zthree$-graded matrices as the ones presented in \cite{akl} can be used as building blocks (see Section {\bf 3})  to construct matrix representations of color Lie (super)algebras. Furthermore,  ternary extensions of color Lie (super)algebras were introduced  in \cite{caraf}.\par
~\par

\par
~\\
{\bf{\Large{Appendix B: on roots of unity and their levels}}}
~\par
~\par
We present here the notions of roots of unity and their level.\par
~\par
Over the field of complex numbers, for any integer $n\in {\mathbb N}$, an $n^{th}$-root of unity denotes any one
of the $n$ distinct roots of the $x^n=1$ algebraic equation.  The level $k$ of a given $n^{th}$-root, where $k$  takes value in the set  $k\in\{1,2,\ldots, n\}$, is determined as follows:
\bea
{\textrm{A root of unity ${\overline x}$ is defined of level-$k$ if $k$ is the minimal positive integer such that
${\overline x}^k=1$.}}&&\nonumber\\&&
\eea
We get, in particular:\\
~\\
$n=1$: $\quad$ $x^1=1\Rightarrow 1~$ solution,   ($x_1=1$, which is a level-$1$ root of unity);\\
$n=2$: $\quad$ $x^2=1\Rightarrow 2~$ solutions ($x_1=1$ and $x_2=-1$, with $x_2$ a level-$2$ root of unity);\\
$n=3$: $\quad$ $x^3=1\Rightarrow 3~$ solutions ($x_1=1$,  $x_2=e^{\frac{2\pi i}{3}}$, $x_3=e^{\frac{4\pi i}{3}}$, with $x_2, x_3$ of level-$3$);\\
$n=4$: $\quad$ $x^4=1\Rightarrow 4~$ solutions  ($x_1=1$,  $x_2=i$, $x_3=-1$, $x_4=- i$ with $x_2, x_4$ of level-$4$)
\bea
{\textrm{and so on for arbitrary values of $n$.}}&&
\eea

An $n^{th}$-root of unity of level $n$ is also called a {\it primitive} root of unity; this term is employed in OEIS - The On-Line Encyclopedia of Integer Sequences, which lists as $A000010$ sequence the number $s_n$ of primitive $n^{th}$ roots of unity.  Up to $n=8$, the $s_1,s_2,s_3,\ldots$ sequence is given by
\bea
A000010: && s_1=1, ~s_2=1,~ s_3=2,~s_4=2,~ s_5= 4,~ s_6=2,~ s_7=6,~ s_8=4, \ldots.
\eea

The physical importance of the notion of ``level-$k$ root of unity" was stressed in \cite{{topqubits},{topvoli}}. It was shown that, 
in the multi-particle quantization of braided Majorana qubits, a level-$k$ root of unity implies a parastatistics such that at most $k-1$ excited states are accommodated in any multi-particle sector. This is a specific implementation of a Gentile-type parastatistics \cite{gen} which extends the Pauli's exclusion principle for ordinary fermions; the ordinary fermionic statistics is recovered from $k=2$. \par
~\\
{\bf Remark}: in certain physical applications, both those presented in \cite{{topqubits},{topvoli}} and the ones discussed in this paper, inequivalent physics is only determined by the root-of-unity level; for each level $k$, each one of its $s_k$ distinct primitive roots produces the same, physically equivalent, results. A corollary of this statement can be seen from the (\ref{z3z3array}) array, where the level-$3$ 
roots of unity $j_1,j_2$ induce isomorphic $\zthreetwo$-graded color Lie algebras.
\\
~\\

{\bf{\Large{Appendix C: on braided Majorana qubits}}}
~\par
~\par
 For selfconsistency of the paper we present here, following \cite{{topqubits},{topvoli}}, the notions of braided Majorana qubits implementing a Gentile-type \cite{gen} parastatistics.\par
A Majorana qubit is introduced, see \cite{topqubits}, as a ${\mathbb Z}_2$-graded qubit expressed by a bosonic  vacuum state $|0\rangle$ and a fermionic excited state $ |1\rangle$ created by the fermionic operator $\gamma$ acting on the vacuum:
{\small{\bea\label{qubit01}
|0\rangle = \left(\begin{array}{c} 1\\0\end{array}\right) , &\quad& |1\rangle =\left(\begin{array}{c} 0\\1\end{array}\right), \qquad{\textrm{with ~~~$\gamma =\left(\begin{array}{cc} 0&0\\1&0\end{array}\right)$  .}}
\eea}}
The reality condition implies that the fermionic excited state coincides with its own antiparticle, i.e. it is a Majorana fermion.\par
The multi-particle braiding is introduced, following the \cite{maj} prescription, in terms of a braided tensor product 
$\otimes_{br}$ satisfying
\bea \label{braidingamma}
({\mathbb I}_2\otimes_{br} \gamma)\cdot (\gamma\otimes_{br} {\mathbb I}_2) &=& B_t\cdot (\gamma\otimes_{br} {\mathbb I}_2)\cdot ({\mathbb I}_2\otimes_{br} \gamma) \equiv B_t\cdot (\gamma\otimes\gamma),
\eea
where the $t$-dependent $4\times 4$ constant matrix $B_t$ is invertible for $t\neq 0$. $B_t$ coincides with the \cite{kasa} $R$-matrix of the Alexander-Conway polynomial in the linear crystal rep on exterior algebra; it
is given by {\footnotesize{\bea\label{btmatrix}
B_t&=&\left(\begin{array}{cccc} 1&0&0&0\\0&1-t&t&0\\0&1&0&0\\0&0&0&-t\end{array}\right).
\eea
}}
 The matrix $B_t$ satisfies the braid relation
\bea\label{braidedrel}
(B_t\otimes {\mathbb I}_2)\cdot ({\mathbb I}_2\otimes B_t)\cdot 
(B_t\otimes {\mathbb I}_2) &=& ({\mathbb I}_2\otimes B_t) \cdot
(B_t\otimes {\mathbb I}_2)\cdot ({\mathbb I}_2\otimes B_t).
\eea
When $t$ is a root of unity ($t= t_k$, labeled by $k=2,3,4,\ldots$)  we have
\bea\label{ts}
(B_{t_k})^k&=&{\mathbb I}_4 \qquad {\textrm{ for \qquad $t_k = e^{\pi i (\frac{2}{k}-1)}$.}}
\eea
As explained in Section {\bf 6}, see formula (\ref{coproducts}), the $N$-particle $\Delta^{(N-1)}(\gamma)$ graded coproduct  of the creation operator $\gamma$, acting on the  $N$-particle vacuum $|0\rangle_N=|0\rangle\otimes |0\rangle\otimes \ldots \otimes |0\rangle$ (taken $N$ times), allows to compute the excited energy levels of the $N$-particle $t_k$-braided Majorana qubits.
The $N$-particle Hamiltonian is $\Delta^{(N-1)}(H)$, where {\footnotesize $H =\left(\begin{array}{cc} 0&0\\0&1\end{array}\right)$}}.\par
The $N$-particle coproduct $\Delta^{(N-1)}(\gamma)$ is expressed by the sum of $N$ nilpotent creation operators
${G}_I^\dagger$, for $I=1,2,\ldots, N$:
\bea\label{nilpotentop}
\Delta^{(N-1)}(\gamma) &=& \sum_{I=1}^N {G}_I^\dagger.
\eea
The nilpotent creation operators satisfy the relations
\bea\label{generalnilpotent}
({G}_I^\dagger)^2 &=& 0 {\textrm{\qquad\qquad \qquad for any $I$, together with}}\nonumber\\
{G}_I^\dagger {G}_J^\dagger &=& s_k\cdot {G}_J^\dagger{G}_I^\dagger {\textrm{\qquad ~for $I<J$, where}}
\nonumber\\
s_k&=& e^{\frac{2\pi i}{k}} \qquad {\textrm{~~ \qquad is a primitive $k^{th}$ root of unity for $k=2,3,\ldots$.}}
\eea
The following combinatorics is easily proved (see \cite{topqubits}):
\bea\label{combinatorics}
\left(\sum_{I=1}^{N} {G}_I^\dagger\right)^n \neq 0 {\textrm{\qquad for ~~~~$n<k$~~~~~~ and}} &~~~&
\left(\sum_{I=1}^{N} {G}_I^\dagger\right)^n = 0 {\textrm{\qquad for ~ $n\geq k$}}.
\eea
This combinatorics has the following implications for the $N$-particle energy levels $E$. 
The levels $E$ are not degenerate and given by integer numbers. In terms of $k$ we get
\bea\label{truncatedenergy}
E &=& 0,1,\ldots, N\qquad \quad~{\textrm{for}}\quad N<k,\nonumber\\
E &=& 0,1,\ldots, k-1 \qquad {\textrm{for}} \quad N\geq k;
\eea
a plateau is reached at the maximal energy level $k-1$, corresponding to the maximal number of braided Majorana fermions that can be accommodated in a multi-particle Hilbert space. \par
This implies that braided Majorana qubits at the given level $k$ implement a Gentile-type \cite{gen} parastatistics
{{(we recall that Gentile extended the notion of fermions by allowing up to $k-1$ excited states to be accommodated in a multi-particle sectors, with ordinary fermions recovered for $k=2$)}}.\par
In the $k\rightarrow\infty$ limit $(t_\infty=-1)$ no plateau is reached and the maximal energy eigenvalues grow linearly with $N$:
\bea\label{genericenergy}
E &=& 0,1,\ldots, N\qquad {\textrm{for any given}}\quad N.
\eea
Further analysis presented in \cite{topvoli} proved two extra features of braided Majorana qubits:\par
~{\it i}) The (\ref{truncatedenergy}) level-$k$ truncations of the energy spectra are recovered by superselected reps, at roots of unity, of the quantum superalgebra  ${\cal U}_q({\mathfrak{osp}}(1|2))$ introduced in \cite{kure} and
\par
{\it ii}) With the help of the $2\times 2$,  level-$k$ intertwiner operator $W_{t_k}$ for $t_k$, the ${\otimes}_{br}$ braided tensor product is expressed as an ordinary tensor product via the positions:
\bea\label{matrixrepof}
 (\gamma~\otimes_{br}{\mathbb I}_2)\mapsto \gamma\otimes {\mathbb I}_2, &\qquad&
 ({\mathbb I}_2~\otimes_{br}\gamma)\mapsto W_{t_k}\otimes \gamma, \qquad\quad {\textrm{so that }}\nonumber\\
 ({\mathbb I}_2~\otimes_{br}\gamma)\cdot 
 (\gamma~\otimes_{br}{\mathbb I}_2)&\mapsto& (W_{t_k}\otimes\gamma)\cdot(\gamma\otimes {\mathbb I}_2)= (W_{t_k}\gamma)\otimes\gamma\nonumber\\
 (\gamma~\otimes_{br}{\mathbb I}_2)\cdot ({\mathbb I}_2~\otimes_{br}\gamma)&\mapsto&
(\gamma\otimes {\mathbb I}_2)\cdot (W_{t_k} \otimes\gamma) = (\gamma W_{t_k})\otimes\gamma.
\eea
At $t\equiv t_k$ the (\ref{braidingamma}) relation is recovered for $W_{t_k}$ satisfying
\bea\label{consistencyintertwining}
W_{t_k}\cdot \gamma &=& (-t_{k}) \gamma \cdot W_{t_k}.
\eea
A solution of this equation, expressed in terms of $k$ entering $t_k$ in (\ref{ts}), is given by  
\bea\label{solutionintertwining}
W_{t_k}&=& \cos(\frac{\pi}{k} )\cdot {\footnotesize{\left(\begin{array}{cc} 1&0\\0&1\end{array}\right) }}- i \sin(\frac{\pi}{k} )\cdot{\footnotesize{ \left(\begin{array}{cc} 1&0\\0&-1\end{array}\right) }}.
\eea
For $N=2$ the identification of the nilpotent operators ${G}_1^\dagger, ~ {G}_2^\dagger$ entering (\ref{nilpotentop}) is
\bea
{G}_1^\dagger = W_{t_k}\otimes \gamma, && 
{G}_2^\dagger = \gamma\otimes {\mathbb I}_2.
\eea

Braided Majorana qubits induce \cite{topvoli} mixed-bracket Heisenberg-Lie superalgebras defined in terms of $\theta_{x,y}$ angles for any pair of $X,Y$ generators. The corresponding round brackets $(\cdot,\cdot)$ are introduced as
\bea\label{roundbracket}
\left(X,Y\right)_{\theta_{x,y}}&:= & i \sin (\theta_{x,y}) [X,Y]+ \cos (\theta_{x,y})\{X,Y\}~= ~e^{i\theta_{x,y}}\left(XY+e^{-2i\theta_{x,y}}YX\right).
\eea
The round brackets of the braided Majorana qubits, labeled by $k$ entering formula (\ref{matrixrepof}), are expressed by the $k$-dependent angles $\vartheta_k$ defined as
\bea
\vartheta_k &=& \frac{k+2}{2k}\pi \qquad{\textrm{for \quad $k=2,3,4,\ldots$.}}
\eea
Some interesting cases are
\bea
k=2&:& \qquad {\textrm{with \quad $\vartheta_{k=2} = \pi$ \quad and \quad $(X,Y)_{\vartheta_{k=2} }= -\{X,Y\}$,}}\nonumber\\
k=3&:& \qquad {\textrm{with \quad  $\vartheta_{k=3} = \frac{5}{6}\pi$,}}\nonumber\\
\ldots &&\nonumber\\
k=6&:& \qquad {\textrm{with \quad  $\vartheta_{k=6} = \frac{2}{3}\pi$,}}\nonumber\\
\ldots &&\nonumber\\
k\rightarrow\infty &:& \qquad {\textrm{with $\vartheta_{k=+\infty} = \frac{\pi}{2}$ \quad and \quad $(X,Y)_{\vartheta_{k=+\infty} }= i[X,Y]$.}}
\eea
~\\
For $2$-particle braided Majorana qubits at level $k$ a $(\cdot,\cdot)$ round-bracket Heisenberg-Lie superalgebra is closed, see \cite{topvoli}, on $5$ generators $G_0, G_{\pm 1}, G_{\pm2 }$. 
One gets, with the identifications
$G_0:={\mathbb I}_4, ~G_{+1}:= G_1^\dagger,~ G_{+2}:= G_2^\dagger,~ G_{-1} := (G_1^\dagger)^\dagger,~ G_{-2}:= (G_2^\dagger)^\dagger$:
\bea\label{slevelmixed}
{\textrm{for $i=1,2$,}} &&
~~ (G_0,G_{\pm i})_{\theta=\pm\frac{\pi}{2}}=(G_{\pm i},G_0)_{\theta=\pm\frac{\pi}{2}}=(G_0,G_0)_{\theta=\frac{\pi}{2}}=0,\nonumber\\
&&~~(G_{\pm i}, G_{\mp i})_{\theta=0} = G_0, \qquad (G_{\pm i}, G_{\pm  i})_{\theta=0}=0,\nonumber\\ 
&&~~(G_{\pm 1}, G_{\pm 2})_{\vartheta_k}~=(G_{\pm 2}, G_{\pm 1})_{-\vartheta_k}=0,\nonumber\\
&&~~(G_{\pm 1}, G_{\mp 2})_{-\vartheta_k}=(G_{\pm 2}, G_{\mp 1})_{\vartheta_k}=0.
\eea

~\par
\par {\Large{\bf Acknowledgments}}
{}~\par{}~\par
We have profited of important discussions and clarifications on various points by several researchers. We mention, in particular, Vera Serganova and Ilya Zakharevich about connections with Volichenko's algebras during the ISQS29 Prague's meeting; the important discussions with the IQOQI 
group (Thomas D. Galley, Manuel Mekonnen and Markus P. M\"uller) on parastatistics, trilinear relations and Young tableaux;  the online discussions with Ruibin Zhang about color Hopf algebras and braiding tensor products; F. T. is grateful to Gerald A. Goldin for clarifications on both higher-dimensional reps of the permutation group and the role of non-abelian anyons. Last, but not least,  we acknowledge the UFABC undergraduate student Jo\~ao Ricardo Rios Mattos who classified the inner products for the ${\mathbb Z}_p\times {\mathbb Z}_q$ groups. His result has not been used in this work; it will be applied to more general color extensions.\par
 The work was supported by CNPq (PQ grant 308846/2021-4).

\end{document}